\documentclass[preprint,11pt,authoryear]{elsarticle}

\usepackage{soul}
\usepackage{amssymb}
\usepackage{lineno,hyperref}
\usepackage{graphicx}
\usepackage{array}
\usepackage{color}
\usepackage{mathptmx}
\usepackage[doublespacing]{setspace}
\usepackage{multirow}
\usepackage{rotating}
\usepackage{algorithm}
\usepackage{algpseudocode}

\newtheorem{theorem}{Theorem}
\newtheorem{corollary}{Corollary}

\usepackage{amsmath}
\newcommand{\argmin}{\mathop{\mathrm{argmin}}\limits}
\newcommand{\argmax}{\mathop{\mathrm{argmax}}\limits}

\journal{Economic Analysis and Policy}

\begin{document}
\begin{frontmatter}

\title{GLOBALIZATION? TRADE WAR? A COUNTERBALANCE PERSPECTIVE
\footnote[4]{Published in \textit{Economic Analysis and Policy} 88 (2025), 1008--1035.\ \url{https://doi.org/10.1016/j.eap.2025.09.029}}
\footnote[2]{The views expressed herein are those of the author and should not be attributed to the IMF, its Executive Board, or its management.}
}

\author[inst]{Xingwei Hu\footnote[3]{The author thanks Reda Cherif, Alexei Goumilevski, Arthur Hu, Jinpeng Ma, Hui Tong, and Wei Zhang for their valuable suggestions, as well as anonymous referees for insightful comments.}}
\address[inst]{International Monetary Fund, Washington, DC 20431, USA; xhu@imf.org}






\begin{abstract}
This paper investigates the evolving dynamics of international trade, emphasizing the strategic interplay between competition and cooperation within the global trade network. It argues that competitive advantages --- rather than traditional comparative advantages --- are the primary drivers of trade conflicts and deglobalization. Drawing on the concept of the balance of power, the paper introduces a quantitative measure of competitiveness, which complements the trade balance as a long-term policy objective. It further explores how countries can enhance competitiveness and trade balance through globalization, protectionism, collaboration, or trade frictions. Using real-world trade data from 2000 to 2019, our empirical study finds parallels between historical developments and quantitative evidence based on this new theory, offering actionable insights to policymakers for managing trade relations, mitigating conflicts, and calibrating the optimal level of globalization.
\end{abstract}


\end{frontmatter}





\section{Introduction}

Adam Smith's concept of economies of scale and David Ricardo's theory of comparative advantage suggest that global production could be maximized if all factors of production moved freely. 
Yet, prosperity remains uneven: some nations flourish while others falter. 
Even historically dominant economies have faced setbacks, underscoring the importance of fair competition. 
To navigate this landscape, nations must strike a balance between competition and cooperation --- securing their fair share while contributing to global output. Both dynamics are essential. 

This paper proposes strategic and fair solutions for countries engaged in international trade.
We identify three key challenges:
\begin{enumerate}
\item {The Dual Nature of Trade:} International trade embodies both conflict and cooperation. 
While voluntary exchange reflects collaboration, a country's growing strength can undermine its partners.
Models that are purely cooperative or non-cooperative fail to capture this complexity.

\item {System-Level Representation:} A framework that emphasizes competitiveness is crucial, as trade flows already reflect comparative advantages.
We conceptualize international trade as a directional network and aim to formulate utility functions that integrate both conflict and cooperation.
Policy variables --- such as imports and exports as shares of production --- should align with these objectives, enabling countries to optimize their strategies.

\item {Data-Driven Insights:}
Real-world data can illuminate past events and guide optimal strategies for the near future --- such as selecting trade partners, resolving disputes, and determining the appropriate degree of globalization.
\end{enumerate}

We model international trade as a network game, where countries are nodes and trade flows are directional edges (e.g., Chaney, 2014; $\ddot{\mathrm{O}}$nder and Yilmazkuday, 2016).
In this multilateral setting, unilateral actions often lead to unpredictable or counterproductive outcomes. 
Trade wars and anti-globalization measures --- though intended to protect domestic industries or reduce deficits --- frequently cause more harm than good (e.g., Chung et al., 2016; Read, 2005). 

A realistic approach requires mutual understanding of each partner's strengths and weaknesses,
as reflected in domestic production and trade flows. These flows, along with domestic consumption, form a coalitional game for each nation.
The Shapley value (1953) of the game provides a revealed preference or individual utility function (Roth, 1977) and serves as a power index in completing domestic production.

The commonly shared utility function --- referred to as the social welfare function --- aggregates individual utility functions (e.g., Harsanyi, 1955; Hammond, 1987) and balances power within the network (Hu and Shapley, 2003).
It evaluates national competitiveness and captures the dual dynamics of cooperation and competition, thereby providing a battleground for both across nations. 

Competitiveness generates additional welfare beyond economies of scale --- a phenomenon often described as the Matthew effect --- and offers further incentive for collaboration, complementing traditional comparative advantages.
While comparative advantage refers to producing goods or services at a lower opportunity cost, thereby promoting  specialization and inclusive growth, competitive advantage emphasizes strategic management aimed at achieving national dominance, fostering exclusive growth.

We adopt a unified game-theoretic framework to analyze two distinct forms of economic conflict: deglobalization and trade wars. 
Deglobalization refers to a country's withdrawal from the global economy, whereas trade wars involve direct confrontation between trading partners.
Both phenomena can significantly disrupt global economic activity. 
This paper proposes clear and practical strategies to mitigate these conflicts.
In our framework, national competitiveness is modeled as a constant-sum game, suggesting that its pursuit often fuels trade wars and anti-globalization measures. In contract, comparative advantage promotes globalization and trade cooperation.

Our approach centers on trade flows, recognizing that geopolitical factors --- such as political, territorial, ideological, cultural, and security considerations --- are inherently embedded in trade data. These data also reflect transaction costs, resource endowments, industrial structures, and locational advantages.
From this perspective, a country's primary objective in trade wars or deglobalization is to improve its trade balance and competitiveness.

The goal of this paper is to offer strategic guidance for policymakers to enhance national competitiveness, rather than focusing on firm-level performance.
Unlike trade firms, which primarily exploit comparative advantages, 
nations are not mere aggregations of firms. Policymakers must also address the adverse effects of trade, such as labor market disruptions and manufacturing offshoring.
Accordingly, this study moves beyond production-based foundations (e.g., Grossman and Helpman, 1995; Harrison and Rutstrom, 1991; Ossa, 2014) and instead seeks to maximize a social welfare function that emphasizes domestic production decisions shaped by global demand.
This function incorporates comparative advantages, weighted by consumer preferences across nations, products, and income levels. 

When a country initiates a trade war, it typically raises tariffs or imposes barriers to boost competitiveness.
The outcome depends on the partner's response.
We find that a simple threshold --- the ratio of competitiveness --- determines whether conflict or cooperation is optimal.
Additionally, responses may benchmark production ratios to avoid trade deficits.
Using these threshold and benchmark ratios, we derive Nash bargaining solutions (1950) for fair responses.
These solutions establish consistent measures of national bargaining power and offer practical methods for resolving bilateral trade frictions.

For countries without explicit confrontation targets, the strategic choice lies between globalization and protectionism.
By adjusting overall exports, we can assess the impact on competitiveness;
the sign of this effect indicates whether deeper globalization is beneficial.
Additionally, our analyses of trade wars, globalization, and bargaining power all help form and reform multilateral trade agreements.

The proposed solutions are data-driven, relying exclusively on trade and production data.
While prior research has examined bilateral or multilateral negotiations under specific assumptions (e.g., Bagwell et al., 2020; Schneider, 2005), and globalization has been extensively debated (e.g., Stiglitz, 2002 and 2017),
these discussions often yield opposing conclusions depending on assumptions and priorities.
In contrast, our data-driven approach sidesteps normative debates on labor markets, supply chains, environmental concerns, and national security --- factors that trade data may already implicitly capture.
Ultimately, whether a country embraces further globalization or protectionism depends on its unique circumstances at a given point of time.

Using UN ComTrade data (2021), we test our theories against major economic events from 2000 to 2019.
Results suggest that protectionism could have benefited the United States and the United Kingdom, but not China, Germany, Japan, or Russia, when competitiveness was the primary goal.
Data from 2017 indicate that trade wars might have helped the USA maintain competitiveness and reduce deficits. We also assess the China--USA trade war in 2018 and 2019,
noting significant spillover effects on third-party economies.
Our impossibility trilemma shows that the USA should reduce imports if it aims to lower national debt by generating trade surpluses while preserving competitiveness.

This study contributes to the trade literature by introducing a strategic, comprehensive framework for analyzing cooperation and competition in global trade. Unlike traditional studies, it avoids unknown parameters, unnecessary assumptions, and excessive mathematization, which can lead to misleading conclusions and policy implications (e.g., Romer, 2015). 

The remainder of this paper is organized as follows.
We introduce the revealed social welfare function in Section~\ref{sect:objective}. 
Section~\ref{sect:trade_war} formulates conditions for a country to initiate a trade war or form a trade partnership to optimize its competitiveness.
Section~\ref{sect:bargaining} derives Nash bargaining solutions for resolving bilateral conflicts and sharing profits.
Section~\ref{sect:globalization} discusses conditions for choosing globalization or protectionism.
Using trade data from 2000 to 2019, Section~\ref{sect:empirical} presents an empirical analysis of historical events. 
Finally, Section~\ref{sect:discussions} concludes with policy implications, extensions, and limitations of the framework.
All proofs are provided in the Appendix, and the exposition is self-contained.

\section{A Network-based Social Welfare Function}\label{sect:objective}

This section introduces a novel concept of competitiveness, which is a social welfare function for countries within the global trade network.
Competitiveness is defined as a relative measure derived from a constant-sum game. 
Unlike absolute production or total trade volume, which provide limited insight into an economy's competitive position, 
this measure captures the strategic dynamics of trade.

The game employs inflows and outflows within the network. In the absence of changes, these flows converge to a static equilibrium over time, reflecting long-term accumulation.
This equilibrium yields a competitiveness vector, which represents the utilities of countries' positions in the network game.
We outline the steps for constructing this data-driven social welfare function and highlight its unique properties, which are largely unfamiliar in the trade literature.

For policymakers, domestic production is a primary concern, whereas firms prioritize profit maximization. 
Production that yields zero or slightly negative profit may be unattractive to firms but can still be a viable policy option for governments seeking to boost employment.
For example, maximizing production can reduce unemployment, while firms might prefer outsourcing labor overseas to maximize shareholder returns. 
Accordingly, our objective function and strategic responses focus on production decision-making, leaving the allocation of production factors and profit distribution to firms. 
Governments, however, play a dual role: regulating profit-maximization behavior and enhancing consumer welfare, while redistributing firm profits through taxation.

To model trade relationships, we represent aggregate trade activities using a square matrix $P$.
Assume there are $n$ countries, labeled $1, 2, \cdots, n$, collectively denoted by the set $\mathcal{N} = \{1,2,\cdots, n\}$.
For any $i,j \in \mathcal{N}$ with $i\neq j$, let $P_{ij}$ denote the fraction of country $i$'s production exported to country $j$.
Exports include all goods and services, including intermediate goods, which may incorporate components from third countries or from country $j$ itself.
The production value of country $i$, denoted $g_i$, consists of all types of exports, all domestically produced final products, and imports directly attributable to production (e.g., raw materials, intermediate goods, crude oil).
However, $g_i$ excludes imported final goods not used in domestic production.
We define $P_{ii}=1-\sum\limits_{j\neq i} P_{ij}\ge 0$ as the non-exporting fraction. 
Clearly, $P_{ij}\ge 0$ for all $i, j \in \mathcal{N}$, and for each $i \in \mathcal{N}$, $\sum\limits_{j=1}^n P_{ij} = 1$.
All fractions $P_{ij}$ form an $n\times n$ stochastic matrix $P = [P_{ij}]$.

Each row of the matrix $P$ defines an individual utility function that breaks down a specific country's production from either a demand-sided or consumption-side perspective.
For any subset $S\subseteq \mathcal{N}$, we define the set function $v_i: 2^{\mathcal{N}} \to [0,1]$ by
\begin{equation}\label{eq:production}
v_i(S) \overset{\mathrm{def}}{=\joinrel=} \sum\limits_{j\in S} P_{ij},
\end{equation}
with the convention $v_i(\emptyset)=0$ for the empty set $\emptyset$.
Thus, $v_i(\cdot)$ is a demand-driven production function for country $i$, and $v_i(S)$ represents the value of production driven by the countries in $S$.

Clearly, $(\mathcal{N},v_i)$ defines a coalitional game, and its Shapley value~(1953) is given by the $i$th row of $P$, i.e., $(P_{i1}, P_{i2},\cdots, P_{in})$.
Roth~(1977) demonstrates that the Shapley value is a von Neumann-Morgenstern utility function~(1953). In this context,  $P_{ij}$ measures an appraisal agent's satisfaction when playing the role of country $j$ in the game of $(\mathcal{N},v_i)$, assuming neutrality toward both ordinary and strategic risks (Roth, 1977).
In the matrix $P$, therefore, rows represent individual utility functions of countries and columns represent the arguments or social states of these functions.

Consequently, countries exhibit cardinal preferences in the game: $P_{ij}>P_{ik}$ implies that country $j$ is preferred to $k$.
However, these preferences vary with $i \in \mathcal{N}$.
A social welfare function aggregates these individual utility functions into a single function, incorporating the diverse value judgments of countries in $\mathcal{N}$.
It evaluates the set of countries as a whole and serves as a common platform for competition and cooperation.

\subsection{Long-Run Influence in the Trade System}
With the normalization conditions $v_i(\mathcal{N})=1$ and $v_i(\emptyset)=0$, and the monotonicity of $v_i$, the measure $P_{ij}$ closely resembles the Shapley-Shubik power index (1954), which quantifies country $j$'s pivotal role in determining the outcome $v_i(\mathcal{N})$.
In essence, $P_{ij}$ represents a form of \textit{pro rata} power when country $i$ realizes its full production. 

The multilinear extension of $v_i$ is defined as:
$$
\tilde v_i (x_1,x_2,\cdots, x_n) \overset{\mathrm{def}}{=\joinrel=} \sum\limits_{S\subseteq \mathcal{N}} \prod\limits_{j \in S} x_j \prod\limits_{k \not \in S} (1-x_k) v_i(S).
$$
Here, $x_j\in [0,1]$ denotes the probability that country $j$ participates in the consumption-driven production.
When restricted to the corners of the unit cube $[0,1]^n$, $\tilde v_i$ simplifies to the original function $v_i$.
Under the assumption of independent actions, $\tilde v_i (x_1,\cdots, x_n)$ represents the expected production of country $i$.

As production evolves from zero at the origin $(0,0,\cdots,0)$ to full completion at $(1,1,\cdots,1)$,
each $x_j$ also reflects country $j$'s progress in completing its imports from country $i$.
The partial derivative of $\tilde v_i$ with respect to $x_j$ captures country $j$'s instant influence or power to increase $\tilde v_i$ at a given point.

In the absence of prior knowledge about participation probabilities or progress, we assume uniform advancement along the diagonal from $(0,\cdots,0)$ to $(1, \cdots,1)$. 
For instance, if country $j$ completes 25\% of its consumption, all other countries are assumed to have completed 25\% as well.
Consequently, the integral of the gradient of $\tilde v_i$, denoted $\bigtriangledown \tilde v_i$, along this diagonal quantifies each country's overall influence in driving country $i$'s production to completion.
According to Owen (1972), this integral equals the Shapley value for the coalitional game:
$$
\int_0^1 \bigtriangledown \tilde v_i (x, \cdots, x) \mathrm{d} x = (P_{i1}, P_{i2},\cdots, P_{in})^\top
$$
where ``$\top$'' denotes vector or matrix transpose.
Therefore, this matrix $P$ contains $n\times n$ quantitative measures of power, and the balance of power in $P$ indicates a stable state in the trade system. 

From a supply-side perspective, the inputs to the production function in Eq.~(\ref{eq:production}) may include any variables that influence the production and services of country $i$.
However, compiling a comprehensive list of these variables is impractical.
Their effects vary over time and across countries, and each product integrates them in unique ways.
For example, inflows of financial assets --- such as remittances and foreign direct investments --- play a significant role  in domestic production.
These assets often represent payments for past, current, or future exports.
Trade involves the exchange of goods and services; and monetary flows facilitate these transactions in modern international trade. 
Yet, payments alone do not fully capture trade activities. Transactions between two firms may be partially offset, resulting in net payments only.
Intra-firm trade within multinational corporations may involve zero payments.
Moreover, trade surpluses and deficits can be influenced by currency manipulation --- depreciation and appreciation, respectively. 

A country may accumulate trade deficits if it is a highly attractive destination for foreign investment. 
For instance, U.S. Treasury Bonds became lucrative and low-risk investment instruments after the Cold War.
As a result, foreign governments or companies often seek to earn additional U.S. dollars through trade to expand their  investment portfolios.

In a closed global trade system, direct influences captured in matrix $P$ naturally give rise to indirect ones .
Within a production chain, even if the final product is labeled as originating from a single country, its components often come from multiple sources.
For instance, when China exports a smartphone to the USA, the chip and design software may be sourced from the USA, the camera from Korea, and the camera screen manufactured in Japan using US patents. 

To account for these spillovers, we consider both indirect and long-term impacts through successive powers of $P$, such as $P^2, P^3, \cdots, P^\infty$.
In general, $P^t$ represents the aggregate $t$-step impacts across all supply chains in the globalized economy. 
Specifically, the entry at the $i$th row and $j$th column of $P^t$, denoted $\left ( P^t \right )_{ij}$, captures all paths of length $t$ from country $i$ to country $j$:
\begin{equation}\label{eq:sumofsum}
\left ( P^t \right )_{ij} 
= 
\sum_{k_1=1}^n \sum_{k_2=1}^n \cdots \sum_{k_{t-1}=1}^n P_{i,k_1} P_{k_1,k_2} P_{k_2,k_3} \cdots P_{k_{t-2},k_{t-1}} P_{k_{t-1},j}.
\end{equation}
Here, each term $P_{k_s,k_{s+1}}$ represents the spillover from country $k_s$ to $k_{s+1}$, or equivalently, the direct influence of $k_{s+1}$ on $k_s$, for $s=1,2,\cdots,t-1$.
These interactions span all countries in the set $\mathcal{N}$.

Under general conditions --- specifically, aperiodicity and irreducibility --- the rows of $P^\infty$ converge to a unique constant row vector $\pi$.
These conditions are naturally satisfied in our trade system.
The element $\pi_j$ is defined as: $\pi_j = \lim\limits_{t\to\infty} \left( P^t \right)_{ij}$ and is independent of the choice of $i$, thereby measuring country $j$'s long-run influence on the entire trade system.
The existence and uniqueness of $\pi$ are well-established in the theory of Markov chains (e.g., Karlin and Taylor, 2012).
Given the vast number of interlinked global value chains, convergence of $P^t$ requires a sufficiently large $t$.
Moreover, the limit of $P^t$ offers a practical computational approach to determine $\pi$.

\subsection{Mixed Cooperation and Noncooperation}
The $1\times n$ row vector $\pi$, referred to as the \textit{authority distribution} in Hu and Shapley (2003), also satisfies the following counterbalance equilibrium: 
\begin{equation} \label{eq:equilibrium}
\pi = \pi P
\end{equation}
subject to the normalization conditions $\sum\limits_{i=1}^n \pi_i = 1$ and non-negativity condition $\pi_i\ge 0$.
Counterbalanced systems are prevalent in both the physical and human domains --- examples include ecological systems, the U.S. government structure, and China's five-element theory.

The counterbalance expressed in Eq.~(\ref{eq:equilibrium}) reflects the mixed cooperative and noncooperative nature of $\pi$.
This duality distinguishes comparative advantage from competitive advantage: while both encourage collaboration,  competitive advantage can also lead to trade wars and deglobalization.
For example, in the trade relationship between China and the United States from 1980 to 2020, the first thirty years were marked by greater cooperation than the last ten.

Although this mixture of cooperation and noncooperation is less familiar to empirical macroeconomics, it may offer valuable insights moving forward (e.g., Allen, 2000, page 147). 
Also, econometric models often struggles to capture the unity of opposites, as estimated coefficients or effects cannot simultaneously be significantly positive and significantly negative.

From a noncooperative standpoint, given that $\sum\limits_{i=1}^n \pi_i = 1$, an increase in $\pi_j$ may imply a decrease in $\pi_i$.
Theoretically, this leads to $\dfrac{n(n-1)}{2}$ potential trade conflicts --- ranging from minor disagreements to major disputes --- within the trade system.
Nonetheless, even minor disputes are worth addressing before they escalate into serious or catastrophic issues.
Moreover, the increase in $\pi_j$ is not uniformly offset by reductions in other economies.
These shifts often follow shared causal patterns triggered by changes in the matrix $P$.
In Sections~\ref{sect:trade_war} and \ref{sect:globalization}, we differentiate between the first move off the diagonal and the move along the diagonal of $P$, corresponding to bilateral trade wars and globalization, respectively.

From a cooperative perspective, country $i$ should assist country $j$ in improving $\pi_j$ whenever $P_{ji}>0$, because $\pi_jP_{ji}$ contributes to $\pi_i$, as shown in:
\begin{equation}\label{eq:power_in}
\pi_i = \sum\limits_{j=1}^n \pi_j P_{ji}
\end{equation}
from Eq.~(\ref{eq:equilibrium}). 
Eq.~(\ref{eq:equilibrium}) also implies that
$
\pi = \pi P = \pi P^2 = \cdots=\pi P^t
$
where:
\begin{equation}\label{eq:pi_P_t}
\pi_i = \sum\limits_{j=1}^n \pi_j \left (P^t \right )_{ji}
\end{equation}
for any $t=1,2,3,\cdots.$
When $\pi_i>0$, as $\lim\limits_{t\to \infty} \left (P^t \right )_{ji} = \pi_i>0$, it follows that $\left (P^t \right )_{ji}>0$ for all sufficiently large $t$ and for all $j\in \mathcal{N}$.
Thus, Eq.~(\ref{eq:pi_P_t}) further implies that country $i$ should assist all other countries in the trade system --- including the poorest and least competitive ones --- to enhance their $\pi_j$.

Therefore, the double-edged counterbalance suggests that a carrot-and-stick approach would better leverage country $i$'s position in the trade system when dealing with its trade partner $j$.
The key trade-off lies in determining how much assistance or contestation to provide to $j$ without sacrificing national interests while potentially increasing $\pi_i$.

\begin{figure}[ht]
\centering
$$
\resizebox{.4\hsize}{!}{$
\begin{array}{ccc}
\fbox{\small $\pi_j=\sum\limits_{i=1}^n \pi_i P_{ij}$}&\overset{\mathrm{exports}}{\xrightarrow{\hspace*{1.5cm}}}&\fbox{\small $\pi_j P_{ji}$}\\
\rotatebox[origin=c]{90}{$\overset{\mathrm{imports}}{\xrightarrow{\hspace*{.9cm}}}$}&&\rotatebox[origin=c]{270}{$\overset{\mathrm{imports}}{\xrightarrow{\hspace*{.9cm}}}$}\\
\fbox{\small $\pi_i P_{ij}$}&\overset{\mathrm{exports}}{\xleftarrow{\hspace*{1.5cm}}}&\fbox{\small $\pi_i=\sum\limits_{j=1}^n \pi_j P_{ji}$}\\
\end{array}
$}
$$
\caption{Dynamics of authority flow for country $i$ and $j$}\label{fig:dynamics_flow}
\end{figure}

We can conceptualize $\pi_i$ as a container or pool within the dynamics of counterbalance equilibrium, where comparative and competitive advantages serve as the primary forces driving the flows.
In the inflow equation Eq.~(\ref{eq:power_in}), country $i$ accumulates or absorbs authority from other countries through its direct influences. 
It gains more authority from influential trade partners (i.e., when $\pi_j$ is large) than from less influential ones, all else being equal.
Additionally, it derives more authority from countries on which it exerts significant direct influence (i.e., when $P_{ji}$ is large), again holding other factors constant. 
As shown in Figure~\ref{fig:dynamics_flow}, country $i$ imports goods from $j$ through the right column, and thus authority flows from $j$ to $i$ along the same column.

Conversely, country $i$ also contributes to other countries, as illustrated in the outflow equation:
$$
\pi_j = \pi_i P_{ij} + \sum\limits_{k\neq i} \pi_k P_{kj}.
$$
The larger $P_{ij}$ or $\pi_i$, the more $i$ contributes to $j$'s authority $\pi_j$, all else being equal. 
In Figure~\ref{fig:dynamics_flow}, country $i$ redistributes $\pi_i$ through the bottom row, according to its export shares. 

These two countries reach a break-even point when:
\begin{equation}\label{eq:pi_i_pi_j_breakeven}
\pi_i P_{ij} = \pi_j P_{ji}
\end{equation}
where $\pi_i P_{ij}$ and $\pi_j P_{ji}$ represent the authority flows from $i$ to $j$ and from $j$ to $i$, respectively.
Intuitively, if $\pi_i P_{ij} < \pi_j P_{ji}$ or $\pi_i P_{ij} > \pi_j P_{ji}$, then country $i$ either gains more authority from or gives more to country $j$, respectively.
For country $i$, a straightforward implication is to decrease or increase its imports from $j$ if $\pi_i P_{ij} < \pi_j P_{ji}$ or $\pi_i P_{ij} > \pi_j P_{ji}$, respectively.
It can also adjust its exports to $j$ accordingly. 

However, the outflow from $i$ to $j$ does not necessarily match the inflow.
With the involvement of third parties, inflows and outflows meet at the equilibrium described by Eq.~(\ref{eq:equilibrium}) for $i$:
$$
\pi_i P_{ij} + \sum\limits_{k\neq j} \pi_i P_{ik} = \pi_j P_{ji} + \sum\limits_{k \neq j} \pi_k P_{ki} = \pi_i.
$$ 
For the same reason, requiring all countries to balance their bilateral trades would be overly restrictive.
For example, country $i$ can still maintain an overall zero net balance if it has a trade surplus with $j$ and a trade deficit with $k$ of the same amount.

Finally, Eq.~(\ref{eq:equilibrium}) implies that $\pi$ is the unit-sum row eigenvector corresponding to the largest row eigenvalue of $P$. 
This represents a form of eigenvector centrality, a subclass of network centrality.  
In the literature, eigenvector centrality is the dominant eigenvector for an adjacency or relation matrix, which often has a zero diagonal (e.g., Bonacich,~1987).
Our matrix $P$ has a nonzero diagonal, allowing domestic and foreign goods to compete in the domestic market and
the same domestic products are used both domestically and overseas.
Hu~(2020) also retains a nonzero diagonal when any accepted student selects only one college to attend from multiple offers; thus, the selected college on the diagonal competes with others that also accept the student and are off the diagonal.
Furthermore, $P$ normalizes the scales of economies such that each row sums to one.
Consequently, $P$ has the largest row eigenvalue $1$, common to all stochastic matrices.

\subsection{$\pi$ as a Measure of Competitive Advantage}\label{subsect:CA}
The competitive advantage of country $i$ refers to its ability to outperform rivals within the global trade system.
For any product exported from country $i$ to country $j$, the competitors of $i$ include all other countries.
Thus, competitive advantage extends from bilateral relations to encompass the entire system --- it reflects the ability to surpass all other countries in market $j$ for any $j\in \mathcal{N}$.
In contrast, comparative advantage is based on an economy's specialization, resource endowment, and technology relative to its trade partner's opportunity cost, making it inherently bilateral.
Globalization, however, may have amplified the importance of competitive advantages over comparative advantages.

According to the literature (e.g., Porter, 1985; Stutz and Warf, 2010), competitive advantage rests on direct control of key production factors, such as access to rare resources, advanced technology, low-cost but highly skilled labor, affordable markets, and inexpensive capital. 
These factors --- whether imported or domestically produced --- are integrated into final products. 
Consequently, imported goods carry the original producer's competitiveness and should be attributed to their source.
For example, in smartphone production mentioned before, competitive factors include U.S. chips and patents, Chinese labor, Japanese precision manufacturing, Korean camera technology.
Various studies also outline qualitative strategies to enhance firm competitiveness.
Porter (1985), for instance, identifies three generic strategies: cost leadership, differentiation, and focus.

We propose that $\pi$ quantifies national competitiveness.
To introduce this quantitative measure, we temporarily ignore home bias in trade (cf., McCallum, 1995) and assume no bilateral trade surplus or deficit.
Denote $\tau_j$ as the $j$th row of $P$.
Under the assumption of no bilateral surplus or deficit, $\tau_j$ --- excluding its $j$th element --- represents imports from other countries into $j$, expressed as fractions of $j$'s production.
This measures other countries' aggregate comparative advantages over $j$.

Moreover, all countries compete for comparative advantages in market $j$.
A large value in $\tau_j$ indicates strong competitiveness in capturing a significant share of $j$'s imports.
If we use market $j$ as the reference point for ranking all countries' competitiveness, the resulting score is the vector $\tau_j$, which sums to one. 
The $j$th element is included because domestic products also compete with imports when consumers make purchasing decisions, assuming no home bias.

We apply endogenous weighting to the reference scores $\tau_1, \tau_2, \cdots, \tau_n$.
These individual cardinal utilities are not directly comparable because operating in a highly competitive market is inherently more challenging than in a less competitive one.
To make them comparable, we introduce weights that are endogenously determined rather than exogenously assigned.
Greater weight should be given to more competitive reference markets.

Let the unknown weights be $\rho = (\rho_1, \rho_2, \cdots, \rho_n)$, where $\sum\limits_{j=1}^n \rho_j = 1$ and $\rho_j\ge 0$ for all $j\in \mathcal{N}$. 
The weighted competitiveness score is:
$$
\sum\limits_{j=1}^n \rho_j \tau_j = \rho P.
$$ 
Both the weighted score and the weight vector quantify national competitiveness.
For consistency, the weighted average $\rho P$ should be a positive multiple of the weight vector $\rho$.
By Theorem \ref{thm:uniqueness_pi}, $\rho$ equals $\pi$.

\begin{theorem}\label{thm:uniqueness_pi}
If $\rho P = c \rho$ for some constant $c>0$, then $c=1$ and therefore $\rho=\pi$.
\end{theorem}

In this endogenous weighting framework, global consumers unconsciously participate in evaluating national competitiveness across all products and services.
Essentially, countries collectively assess themselves without external influence from business interests, media, advertising, governments, or military pressure.

The structure of the global trade network is largely shaped by the distribution of firm productivity within individual countries.
However, government policies can influence this distribution --- without significantly affecting international trade welfare --- by shifting imports or exports between countries (e.g., Gopinath et al., 2024).
Policies can also enhance firm productivity through tax exemptions, subsidies, and imposing tariffs on foreign products.

Importantly, national competitiveness $\pi_i$ is not simply the sum of domestic firms' competitiveness.
Instead, firm competitiveness is weighted in $\pi$ by consumer preferences --- wealthy or poor, domestic or overseas.
Affordability thus plays a critical role in determining $\pi$.
Positive weights imply a positive correlation between firm-level and national competitiveness.

Moreover, $\pi$ represents a von Neumann-Morgenstern Bergson social welfare function, as it is a positively weighted sum of individual von Neumann-Morgenstern utility functions $\tau_1, \tau_2, \cdots, \tau_n$.
This weighted utility function $\pi$ is determined not by factor consumption or added value, but by consumer preferences, independent of national interests.
Pareto optimality is achieved because: 
$$
\pi_j \ge \pi_k\ \mathrm{if}\ P_{ij}\ge P_{ik}\ \mathrm{for}\ \mathrm{all}\ i\in \mathcal{N}.
$$
This optimality remains invariant if any individual utility function is scaled or shifted by a positive constant (e.g., scaling $\tau_i$ by $g_i$).
Harsanyi (1955) and Hammond (1987) provide further ethical, risk, and utilitarian interpretations of this type of social welfare function.

This social welfare function $\pi$ differs significantly from widely used utility or welfare functions in the literature on trade wars and globalization (e.g., Grossman and Helpman, 1995; Harrison and Rutstrom, 1991; Ossa, 2014).
Unlike traditional approaches that specify a functional form based on production factors and total factor productivity, our formulation is implicitly derived from the counterbalance equilibrium, accounting for the indirect effects of imports on final goods and services.
Because exported goods and services yield different utilities for exporting and importing countries, 
each country possesses a distinct utility function. Consequently, a single utility function cannot capture the heterogeneity across nations.

Our approach minimizes assumptions by disregarding how efficiently countries transform production factors into final products or how they engage in trade.
The index $\pi$ is constructed directly from observed data without relying on parametric models.
As a result, there are no unknown parameters to estimate and no residuals to reconcile with a model specification.

Furthermore, as Barney and Felin (2013) note, the debate over micro-foundations remain unresolved, with scholars in management, strategy, and organization offering divergent views on the relationship between macroeconomic phenomena and micro-level behaviors.
Given the vast heterogeneity among firms and nations, modeling micro-founded strategic interactions within a network comprising millions of product flows across tens of thousands of directional edges is particularly challenging.
A certain degree of abstraction is therefore necessary to capture the essential features of such a complex system.

The index $\pi_i$ includes a unique component, $\pi_i P_{ii}$, which originates entirely from domestic sources, as indicated in Eq.~(\ref{eq:power_in}).
This component encompasses critical resources such as oil fields, natural gas reserves, rare minerals, and patented technologies --- factors that remain central to a nation's competitive advantage.

A global value chain typically consists of a finite sequence of steps but also includes indefinite domestic loops when tracing the value added to final products.
For example, in the case of smartphones, the education, skills, and experience of manufacturing labor are developed domestically over many years, contributing to the final product.
Similarly, U.S. workers invest significant time in developing patents and design software domestically.
These internal loops influence other countries but are not reciprocally affected, underscoring the role of domestic competitiveness $\pi_i P_{ii}$.
Elements such as social infrastructure, political institutions, monetary and fiscal policies, and the quality of the microeconomic environment all contribute to national competitiveness (e.g., Delgado et al., 2012).

Within a specific global value chain, the opposite end of this unique component involves the design and assembly of the final product. 
Product leadership integrates market orientation, consumer preferences, strategic direction, technology innovation, sales, and production factors --- representing another dimension of competitiveness. 
Between raw resources and the final product lies in a series of value-added processes for intermediate goods,
where competitiveness is driven by cost efficiency, including financing costs.
Technological progress reduces labor costs, while public safety and social stability mitigate risk-related expenses.

Eq.~(\ref{eq:sumofsum}) reveals a strong positive correlation between $\pi_i$ and $g_i$, which reflect the depth and breadth, respectively, of a country's economy.
After adjusting for the magnitude of $g_i$ in matrix $P$, the interaction between competitiveness and economies of scale persists, offering notable cost advantages.
Countries tend to export goods they possess in abundance and import those they lack.
When these are significant, country $i$ will exhibit substantial nonzero elements in both the $i$th column and row of matrix $P$, contributing meaningfully to the summation in Eq.~(\ref{eq:sumofsum}).

Large economies also foster a high degree of product and service differentiation, which provides location-specific advantages and reduces transaction costs for domestic consumers.
These goods, being exempt from tariffs and international shipping fees, enjoy a price advantage.
As a result, $P_{ii}$ is often the largest in the $i$th row of $P$, regardless of any home bias.
Therefore, a large and comprehensively integrated industrial system --- such as those found in China or the United States --- is likely, though not guaranteed, to exhibit a high value in $\pi$. 
In contrast, countries with narrow industrial structures are more vulnerable to competitive pressures, and their $\pi_i$ tend to be fragile and volatile, especially when competing with the U.S. in terms of competitiveness.

We thus adopt $\pi_i$ as a strategic objective for country $i$ in the formulation of trade policies. 
It encapsulates both the depth and breadth of the economy, reflecting cooperative and non-cooperative dynamics with other nations.
This metric is endogenously derived from the comparative advantages across all goods and services.
Recognizing that each country possesses unique geographic, natural, human, and institutional characteristics,
we argue that a single utility function cannot adequately represent the diversity of global economies.
Our study embraces the distinctiveness of each nation and its trade relationships, emphasizing the inclusive value each contributes to the global economy --- where international trade remains a foundational pillar.

\subsection{The Matthew Effect in $\pi$}\label{subsect:Matthew}

In Eq.~(\ref{eq:power_in}), country $i$'s competitiveness $\pi_i$ accumulates in proportion to its initial level of competitiveness: 
\begin{equation}\label{eq:initial_proportion}
\pi_i = \pi_i P_{ii} + \sum\limits_{j\neq i} \pi_j P_{ji},
\end{equation}
where the self-reinforcing proportion $P_{ii}$ is typically greater than $.7$. 
This formulation illustrates the Matthew effect --- the phenomenon of accumulated advantage --- often summarized by the aphorism: ``the rich get richer and the poor get poorer.''

For example, the United States may exhibit a substantially higher $\pi_i$ than Japan, compared to their production ratio. 
This amplification arises not only from the U.S.'s internal advantages but also from its relative superiority over other countries compared to Japan.
Specifically, in Eq.~(\ref{eq:initial_proportion}), for $i=\mathrm{USA}$ or $i=\mathrm{Japan}$:
$$
\left\{
\begin{array}{lcl}
\pi_i^{\mathrm{USA}}  &=&\frac{1}{1-P_{ii}^{\mathrm{USA}}} \sum\limits_{j\neq \mathrm{USA}} \pi_j P_{ji}^{\mathrm{USA}}, \\
\pi_i^{\mathrm{Japan}}&=&\frac{1}{1-P_{ii}^{\mathrm{Japan}}} \sum\limits_{j\neq \mathrm{Japan}} \pi_j P_{ji}^{\mathrm{Japan}}. 
\end{array}
\right.
$$
The Mathew effect in $\pi_i^{\mathrm{USA}}$ arises from two inequalities: $1>P_{ii}^{\mathrm{USA}}>P_{ii}^{\mathrm{Japan}}>0$; and $P_{ji}^{\mathrm{USA}} > P_{ji}^{\mathrm{Japan}}>0$, generally for other countries $j$.

Alternatively, the Matthew effect can be interpreted through preferential attachment in endogenous weighting, where  weights are distributed proportionally to existing competitiveness.
This implies that $\pi_i / \pi_j > g_i / g_j$ when $g_i$ is significantly larger than $g_j$. 
Normalizing $g_i$ as $\tilde g_i \overset{\mathrm{def}}{=\joinrel=} \frac{g_i}{\sum_{j=1}^n g_j}$,  the effect suggests  that $\pi_i$ positively correlates with $\tilde g_i^2$, after controlling for $\tilde g_i$.
Thus, $\pi_i$ exhibits quadratic growth as $\tilde g_i$ increases linearly.
In a linear regression framework: 
\begin{equation}\label{eq:pi_g_gsq}
\pi_i = c_1 \tilde g_i + c_2 \tilde g_i^2 + \epsilon_i,
\end{equation}
for some $c_1>0, c_2>0$, and $\epsilon_i$ is the residual term.
Therefore, when country $i$ is rapidly ascending, country $j$ should seek to resolve conflicts with $i$ before $i$  becomes too powerful to compromise.

If country $i$ leverages its accelerating $\pi_i$ to boost production $g_i$, such as:
$$
\tilde g_i = c_3 + c_4 \pi_i + \epsilon_i
$$
with $c_4>0$, then $\tilde g_i$ accelerates quadratically, leading to quartic growth in $\pi_i$.
Iterating this process results in exponential growth of both $\pi_i$ and $\tilde g_i$ toward their maximum capacities. 
Historically, many empires have emerged rapidly, gaining prominence with astonishing speed.

Moreover, the welfare associated with competitive advantage induces an agglomeration effect, clustering countries around a few economic superpowers.
According to Corollary~\ref{cr:agglomeration}, imports from a large economy are preferred over those from a smaller one, all else being equal. 
Consequently, smaller economies may counteract this preference by lowering export prices.
A third party might impose differentiated tariffs on large and small countries to equalize import prices.
Additionally, large markets --- characterized by numerous producers and consumers without a dominant entity --- may be favored for their market competitiveness.
In either case, smaller countries often fear the rise of neighboring superpowers with similar geographic and resource advantages.

\begin{corollary}\label{cr:agglomeration}
Assume the Matthew effect. If $g_j$ is significantly greater than $g_k$, then third party $i$ prefers imports from country $j$ over country $k$, all else being equal.
\end{corollary}

Thus, the additional welfare granted to competitive economies becomes a major force driving trade flows, alongside comparative advantage.
However, trade wars and protectionism are more likely to stem from competitive advantages, which benefit only select partners, rather than comparative advantages, which benefit all.
Accordingly, an economic superpower will strive to preserve its status when challenged by emerging competitors. Maintaining its advantage in $\pi$ helps prevent a reverse Matthew effect.
As Copeland (2014) notes: ``When expectations [of the future trade environment] turn negative, leaders are likely to fear a loss of access to raw materials and markets, giving them an incentive to initiate crises to protect their commercial interests.''
Hence, the incentive is more proportional to $\pi_i$ than to $g_i$.
Emerging superpowers seeking their rightful place face formidable challenges.
World Wars I and II were intense confrontations between established and rising powers, ultimately reshaping the global order.

\section{Bilateral Trade War for Competitiveness}\label{sect:trade_war}

Anticipating a significant positive shift in its competitiveness $\pi_i$, country $i$ may consider initiating a trade war against its trade partner $j$. 
By strategically manipulating the trade matrix $P$, country $i$'s initial move could alter the element $P_{ji}$ --- the share of country $j$'s production exported to country $i$.
For example, increasing tariffs or reducing import quotas on goods from country $j$ would lower imports from $j$, thereby decreasing $P_{ji}$.
Country $i$ might also restrict certain exports to $j$, directly reducing $P_{ij}$. 

For simplicity, we assume the bilateral conflict affects only the two counterparties in matrix $P$, without directly impacting $P_{ik}, P_{ki}, P_{jk}$, or $P_{kj}$ for any third party $k$.
However, changes in $P_{ji}$ or $P_{ij}$ can propagate through global value chains, indirectly involving third parties over multiple steps.
This simplification does not fully capture the complexity of events such as the 2018 China--USA trade war, during which  the Biden administration encouraged allies to prohibit Chinese firms from acquiring advanced chips and chip-making equipment (New York Times, 2022).

\subsection{Reciprocal Retaliations}
In response, country $j$ may retaliate by reducing $P_{ij}$ or $P_{ji}$.
For instance, following the Trump administration's imposition of tariffs and trade barriers on China in January 2018,  
China retaliated in April 2018 by levying tariffs on 128 U.S. products (Washington Post, 2018a).
If country $j$ does not retaliate, the conflict ends at its inception, and country $i$ emerges victorious.
A lack of response or unconditional surrender may prevent further damage but could also invite additional sanctions from country $i$ if country $j$ appears weak.
Conversely, an overreaction to hostile changes in $P_{ji}$ or $P_{ij}$ could also severely harm $\pi_j$.

To formalize retaliation, we introduce a matrix $\Lambda$ representing retaliation actions.
In response to a change $\Delta P_{ji}$ in $P_{ji}$, country $j$ adjusts $P_{ij}$ by $\lambda_{ji} \Delta P_{ji}$, i.e., $\Delta P_{ij} = \lambda_{ji} \Delta P_{ji}$.
Similarly, in retaliation for a change $\Delta P_{ij}$, country $j$ modifies $P_{ji}$ by $\lambda_{ij} \Delta P_{ij}$, i.e., $\Delta P_{ji} = \lambda_{ij} \Delta P_{ij}$.
Thus, $\lambda_{ij} = 1/\lambda_{ji}$, assuming no first-mover advantage.

We assume $\lambda_{ji}>0$ to ensure aligned actions from both countries.
Ignoring temporally lagged reactions on the long-run effects $\pi$,
changing $P_{ji}$ by $\Delta P_{ji}$ and $P_{ij}$ by $\lambda_{ji} \Delta P_{ji}$ yields equivalent impacts.
Letting $\lambda_{ii}=1$, we compile all $\lambda_{ij}$ into an $n\times n$ matrix $\Lambda=[\lambda_{ij}]$, for all $i,j\in \mathcal{N}$.
To preserve the unit sum in each row of $P$, we deduct $P_{jj}$ by $\Delta P_{ji}$ and $P_{ii}$ by $\Delta P_{ij}$.
The reduced exports are then consumed domestically, ensuring that the modified $P$ remains a stochastic matrix.

The coefficient $\lambda_{ji}$ often depends on various exogenous factors from both countries, including political considerations, anti-dumping measures, trade deficits, and the nature of traded commodities or services.
In practice, there is no universal formula for every contingency, and negotiations may involve multiple rounds of bilateral discussions.
In the subsequent section, we assume $\lambda_{ji}$ is an exogenous or predetermined constant in a potential trade war. 
Section~\ref{sect:bargaining} explores a generic yet simple bargaining solution for $\lambda_{ji}$ across most of the $n(n-1)/2$ bilateral conflicts.
The solution aims not to eliminate conflicts but to mitigate them and fairly distribute the associated costs, serving the diverse interests of the involved countries.
In a mixed cooperative and noncooperative setting, simply maximizing $\lambda_{ji}$ or $\lambda_{ij}$ may be of limited interest to either party.

We introduce several notations for the next five theorems and their corollaries.
Let $I_n$ denote the $n\times n$ identity matrix and $\pi_{_{-i}}$ represent the transpose of $\pi$ with its $i$th element removed.
Define $Z_i$ as the transpose of $P$ with its $i$th row and column removed.
The column vector $\alpha_i$ is obtained by extracting the $i$th row from $P$ and removing its $i$th element.
Vectors $\vec{1}_n$ and $\vec{0}_n$ are $n\times 1$ column vectors of all ones and all zeros, respectively.
For any $j=1,\cdots,n$, the $n\times 1$ vector $e_j$ has a $1$ in its $j$th position and zeros elsewhere.
Finally, the vector $\gamma_{ji}$ is derived from $e_j$ by removing its $i$th element.
Thus, $\alpha_i$, $\pi_{_{-i}}$, and $\gamma_{ji}$ are all non-negative $(n-1)\times 1$ column vectors.

\begin{theorem}\label{thm:trade_war}
Under the above setting and notations, for any $j\neq i$:
\begin{equation}\label{eq:pi_i_2_P_ji}
\hspace{.3cm} \frac{\mathrm{d} \pi_i}{\mathrm{d} P_{ji}}
= 
-\frac{(\lambda_{ji}\pi_i-\pi_j)\vec{1}^\top_{n-1} (I_{n-1} -Z_i)^{-1} \gamma_{ji}}{1+\vec{1}_{n-1}^\top (I_{n-1}-Z_i)^{-1} \alpha_i},
\end{equation}
\begin{equation}\label{eq:pi_j_2_P_ji}
\frac{\mathrm{d} \pi_j}{\mathrm{d} P_{ji}} 
= 
\frac{(\lambda_{ji}\pi_i-\pi_j)\vec{1}^\top_{n-1} (I_{n-1} -Z_j)^{-1} \gamma_{ij}}{1+\vec{1}_{n-1}^\top (I_{n-1}-Z_j)^{-1} \alpha_j}, 
\end{equation}
and
\begin{equation}\label{eq:pi_minus_i_2_P_ji}
\frac{\mathrm{d} \pi_{_{-i}}}{\mathrm{d} P_{ji}} 
=
(\lambda_{ji}\pi_i-\pi_j)
\left (I_{n-1}-Z_i \right )^{-1} \left [ 
\gamma_{ji}
-
\frac{\vec{1}^\top_{n-1} (I_{n-1} -Z_i)^{-1} \gamma_{ji}}{1+\vec{1}_{n-1}^\top (I_{n-1}-Z_i)^{-1} \alpha_i}
\alpha_i
\right ].
\end{equation}
\end{theorem}

Using the relation $\mathrm{d} P_{ij} = \lambda_{ji} \mathrm{d} P_{ji}$ and $\lambda_{ij}={1}/{\lambda_{ji}}$, we also have: 
$$
\frac{\mathrm{d} \pi_i}{\mathrm{d} P_{ij}} = \frac{\mathrm{d} \pi_i}{\lambda_{ji} \mathrm{d} P_{ji}} = \lambda_{ij} \frac{\mathrm{d} \pi_i}{\mathrm{d} P_{ji}},\qquad
\frac{\mathrm{d} \pi_j}{\mathrm{d} P_{ij}} = \lambda_{ij} \frac{\mathrm{d} \pi_j}{\mathrm{d} P_{ji}}, \qquad
\frac{\mathrm{d} \pi_{_{-i}}}{\mathrm{d} P_{ij}} = \lambda_{ij} \frac{\mathrm{d} \pi_{_{-i}}}{\mathrm{d} P_{ji}}.
$$ 

Theorem~\ref{thm:trade_war} implies Corollary~\ref{cor:zero_d_pi}, a variant of Eq.~(\ref{eq:pi_i_pi_j_breakeven}) --- $\pi_i$ remains unchanged when the authority flow from $i$ to $j$ equals that from $j$ to $i$.
However, Eq.~(\ref{eq:pi_i_pi_j_breakeven}) does not rely on the assumptions in Theorem~\ref{thm:trade_war}.

\begin{corollary}\label{cor:zero_d_pi}
When $\lambda_{ji}=\pi_j/\pi_i$ (or equivalently, $\lambda_{ij}=\pi_i/\pi_j$),
$\frac{\mathrm{d} \pi}{\mathrm{d} P_{ji}}=
\frac{\mathrm{d} \pi}{\mathrm{d} P_{ij}} = \vec{0}_n^\top$.
\end{corollary} 

Additionally, since 
$
(I_{n-1}-Z_i)^{-1} = I_{n-1}+Z_i + Z_i^2 + Z_i^3 +\cdots
$ 
has all non-negative elements, the derivatives $\frac{\mathrm{d} \pi_i}{\mathrm{d} P_{ji}}$ and $\frac{\mathrm{d} \pi_j}{\mathrm{d} P_{ji}}$ have opposite signs, as stated in Corollary~\ref{cor:opposite_sign}.
In a trade friction (i.e., $\mathrm{d} P_{ji}<0$), $j$'s overreaction with $\lambda_{ji}>\pi_j/\pi_i$ benefits $\pi_i$ but deteriorates $\pi_j$, according to Eqs.~(\ref{eq:pi_i_2_P_ji})--(\ref{eq:pi_j_2_P_ji}).

\begin{corollary}\label{cor:opposite_sign}
For any $i\neq j$,
$
\frac{\mathrm{d} \pi_i}{\mathrm{d} P_{ji}} \frac{\mathrm{d} \pi_j}{\mathrm{d} P_{ji}}
\le 0
$
and
$
\frac{\mathrm{d} \pi_i}{\mathrm{d} P_{ij}} \frac{\mathrm{d} \pi_j}{\mathrm{d} P_{ij}}
\le 0.
$
\end{corollary}

\subsection{Identification of Competitors and Collaborators}\label{sect:identify_competitor_collaborator}
By varying country $j$ in Eq.~(\ref{eq:pi_i_2_P_ji}), we can identify country $i$'s trade collaborators and competitors.
In Eq.~(\ref{eq:pi_i_2_P_ji}), the derivative $\frac{\mathrm{d} \pi_i}{\mathrm{d} P_{ji}}$ has the opposite sign of $\lambda_{ji}\pi_i-\pi_j$.
Accordingly, if $\lambda_{ji} > \pi_j/\pi_i$, then country $j$ becomes a potential target for conflict with country $i$ because a small negative change in $P_{ji}$ would result in a positive change in $ \pi_i$.
Conversely, if $\lambda_{ji} < \pi_j/\pi_i$, then country $j$ is a potential candidate for deeper collaboration.
In this case, when $\lambda_{ji} < \pi_j/\pi_i$ and $\mathrm{d} P_{ji}>0$, a larger $\lambda_{ji}$ slows the growth of $\pi_i$ and accelerates the decline of $\pi_j$.
Finally, if country $j$ maintains $\lambda_{ji}=\pi_j / \pi_i$, then any small change in $P_{ji}$ will have negligible effects on both $\pi_i$ and $\pi_j$, regardless of the direction of $\mathrm{d} P_{ji}$.

To facilitate cross-country comparisons, we express changes in terms of percentage variation in both $\pi_i$ and $P_{ji}$. 
The limit of the percentage change of $\pi_i$ with respect to that in $P_{ji}$ is given by:
\begin{equation}\label{eq:percentchange}
\frac{\mathrm{d} log(\pi_i)}{\mathrm{d} log(P_{ji})}
=
\frac{\mathrm{d} \pi_i / \pi_i}{\mathrm{d} P_{ji} / P_{ji}}
=
\frac{P_{ji}}{\pi_i} \frac{\mathrm{d} \pi_i}{\mathrm{d} P_{ji}}.
\end{equation}

When Eq.~(\ref{eq:percentchange}) exceeds a positive threshold $\theta$ (e.g., $\theta=.1$) or falls below $-\theta$, the derivative in Eq.~(\ref{eq:pi_i_2_P_ji}) is considered significantly positive or negative, respectively.
For instance, if Eq.~(\ref{eq:percentchange}) equals $.2$, then a $1\%$ increase in $P_{ji}$ would approximately lead to a $.2\%$ increase in $\pi_i$.
Therefore, a significantly negative (or positive) derivative in Eq.~(\ref{eq:pi_i_2_P_ji}) serves as a necessary --- but not sufficient --- condition for country $i$ to initiate a trade war (or form an economic alliance) with country $j$, even if Eq.~(\ref{eq:pi_j_2_P_ji}) remains insignificant.

Furthermore, country $i$ would select
\begin{equation}\label{eq:best_partner}
\argmax_{j\neq i} \frac{\mathrm{d} log(\pi_i)}{\mathrm{d} log(P_{ji})}
\end{equation}
as its best trade partner and
\begin{equation}\label{eq:best_cpmpetitor}
\argmin_{j\neq i} \frac{\mathrm{d} log(\pi_i)}{\mathrm{d} log(P_{ji})}
\end{equation}
as its worst competitor.
In Eqs.~(\ref{eq:best_partner}) and (\ref{eq:best_cpmpetitor}), $\pi_i$ experiences the largest increase or decrease, respectively, in percentage terms for each percentage rise in $P_{ji}$.
However, according to Corollary~\ref{cor:opposite_sign}, country $i$ is not necessarily its best partner's best partner, nor its worst competitor's worst competitor.
This implies that international trade dynamics are not akin to sports-like competitions or zero-sum games --- unless the countries involved are evenly matched.
For example, during the 2018 China-USA trade war, both the Trump and Biden administrations actively challenged China, while China largely adopted a defensive posture.

Eq.~(\ref{eq:pi_minus_i_2_P_ji}) describes the spillover effects on other countries resulting from country $i$'s change in $P_{ji}$.
When $P_{ji}$ is perturbed, some countries gain competitiveness while others lose it; these gains or losses can be more substantial than $\frac{\mathrm{d}\pi_j}{\mathrm{d}P_{ji}}$.
Although $\pi_i$ and $\pi_j$ always move in opposite directions, their magnitude differ:
one change may be significant while the other remains negligible, or both may be minor.
Third countries share the difference between these movements unevenly, and their individual or collective impacts can be substantial.

These spillover effects may influence the determination of $\lambda_{ji}$ when country $i$ is significantly more powerful than country $j$.
The conflict appears resolved if the first-order condition $\lambda_{ji} = \pi_j/\pi_i$ maximizes $\pi_i$.
However, if this condition instead maximizes $\pi_j$, then, in theory, $i$ could aggressively counteract.
In practice, such extreme actions are unlikely because $i$'s gain from $j$ is capped by $\pi_j P_{ji}$ (as shown in Eq.~(\ref{eq:power_in})), which may be negligible for $i$.
Moreover, $\pi_j P_{ji}$ could be smaller than the losses incurred by third parties, as measured by Eq.~(\ref{eq:pi_minus_i_2_P_ji}).
Excessively non-cooperative behavior could provoke backlash from these countries, potentially leading $j$ to form alliances and pursue collective bargaining against $i$. 

Alternatively, $j$ might adopt objectives beyond maximizing $\pi_j$, such as avoiding direct confrontation with $i$.
For instance, it could deepen economic engagement with $i$ by developing complementary industries and leveraging its comparative advantages.
Historically, norms and standards have evolved to restrain unilateral trade dominance; today, the World Trade Organization (WTO) plays a central role in regulating trade rules among nations.

\section{Nash Bargaining Solutions for Reprisal Coefficients $\Lambda$}\label{sect:bargaining}

When additional factors are considered, the condition $\lambda_{ji}>\pi_j/\pi_i$ alone is insufficient for country $i$ to initiate a conflict against country $j$.
Among these factors, trade deficits are particularly significant.
This section explores bilateral bargaining solutions for $\lambda_{ji}$ when trade balance is also an objective.

\subsection{Trade Balance}
For any given period, a trade surplus for country $i$ with country $j$ implies an equivalent trade deficit for $j$ with $i$, and vice versa.
A trade surplus creates job opportunities, reduces unemployment, and expands economies of scale. 
It also enhances creditworthiness by enabling debt repayment.
Surpluses often result from comparative and competitive advantages in tradable goods and services.
Opinions on trade deficits vary. While large deficits can undermine economic sustainability,
in the short term, they may help avoid shortages and mitigate issues like inflation and poverty.
The implications of trade deficits also depend on their impact on national security and how they are financed.
No nation can completely ignore trade deficits or competitiveness; a one-sided emphasis is unsustainable.

There are two benchmark choices for $\lambda_{ji}$ regarding net trade balances: $g_j/g_i$ and $P_{ij}/P_{ji}$, as stated in Theorem~\ref{thm:zero_trade_deficit}.
For country $i$ to have a trade surplus with country $j$ due to a change $\Delta P_{ij}$, 
$i$ must export more to $j$ than it imports, i.e., $g_i \Delta P_{ij} > g_j \Delta P_{ji}$.
Thus, $\lambda_{ji} > g_j/g_i$ implies $i$'s trade surplus if $\Delta P_{ji}>0$. 
To sustain trade peace under $\Delta P_{ji}>0$, a large $\lambda_{ji}$ improves $i$'s trade balance but harms $\pi_i$ --- excessive exports from $i$ to $j$ transfer power to $j$, reducing $\pi_i$.
Therefore, $\lambda_{ji}$ should be neither too large nor too small.
Conversely, $\lambda_{ji} < g_j/g_i$ also indicates $i$'s surplus if $\Delta P_{ji}<0$.
In this case, both $i$ and $j$ lose exports, though $i$ loses less. 
Such outcomes can harm future production, so most countries avoid mutually damaging stalemates.
At a cumulative level, $g_i P_{ij}$ and $g_j P_{ji}$ represent exports from $i$ to $j$ and from $j$ to $i$, respectively.
Thus, $g_j/g_i= P_{ij}/P_{ji}$ results in zero trade surplus and deficit over the accumulation period.
During this period, $g_i$, $g_j$, $P_{ij}$, and $P_{ji}$ all grow from zero, so
$\Delta P_{ji} = P_{ji} - 0$, $\Delta P_{ij} = P_{ij} - 0$, and 
$\lambda_{ji} = \dfrac{\Delta P_{ij}}{\Delta P_{ji}} = \dfrac{P_{ij}}{P_{ji}} = \dfrac{g_j}{g_i}$.

\begin{theorem}\label{thm:zero_trade_deficit}
$\lambda_{ji} = g_j/g_i$ or $g_j/g_i= P_{ij}/P_{ji}$ maintains zero net trade deficit and surplus between countries $i$ and $j$ at the instantaneous or cumulative level, respectively.
\end{theorem}

To ease trade tensions, we consider two approaches for determining $\lambda_{ji}$:
if both sides aim for higher competitiveness, then $\lambda_{ji}=\pi_j/\pi_i$ is the steady-state solution for any minor change in $P_{ji}$;
if they seek balanced trade, then $\lambda_{ji}=g_j/g_i$ eliminates future trade deficits.
In general, $\lambda_{ji}=\pi_j/\pi_i$ applies to long-term rivalry because $\pi$ reflects competitiveness across all countries and industries, whereas
$\lambda_{ji}=g_j/g_i$ can immediately resolve trade deficits and alleviate related issues such as unemployment and national debt. 

For example, according to Pozsar (2022), the 2018 China-USA trade war began in this manner: China had grown exponentially over four decades by exporting inexpensive goods overseas, while the USA lost much of its manufacturing base and accumulated massive national debt through wars on terror, significant tax cuts, and the 2007 subprime mortgage crisis.
When China sought to build a global 5G network and produce cutting-edge computer chips, the USA attempted to block these moves and urged its allies to join the effort.
In this case, the focal point was competitiveness rather than trade balance,
suggesting a prolonged conflict where $\pi_j/\pi_i$ is the appropriate choice for $\lambda_{ji}$.

We examine a scenario in which a dominant superpower, country $i$, faces imminent challenges from an emerging economy, country $j$.
According to the Matthew effect, it is likely that $\frac{\pi_j}{\pi_i} < \frac{g_j}{g_i}$.
Based on Theorem~\ref{thm:impossibility_trilemma}, country $i$ cannot simultaneously achieve three objectives in its bilateral trade with country $j$: increasing its competitiveness $\pi_i$; maintaining a trade surplus; expanding exports to country $j$.
This constraint creates what we call the Impossibility Trilemma.
The trilemma implies that if country $i$ attempts to grow exports to country $j$ while increasing competitiveness, it cannot also sustain a trade surplus.
This situation occurs when $\Delta P_{ij}<0$, causing both countries to lose comparative advantages.

\begin{theorem}[Impossibility Trilemma]\label{thm:impossibility_trilemma}
Given $\frac{\pi_j}{\pi_i} < \frac{g_j}{g_i}$, country $i$ cannot simultaneously achieve the three objectives in its bilateral trade with country $j$:
increasing $\pi_i$; a trade surplus; positive export growth (i.e., $\Delta P_{ij}>0$).
\end{theorem}

\begin{figure}[ht]
\centering
\includegraphics[height=5.3cm, width=9cm]{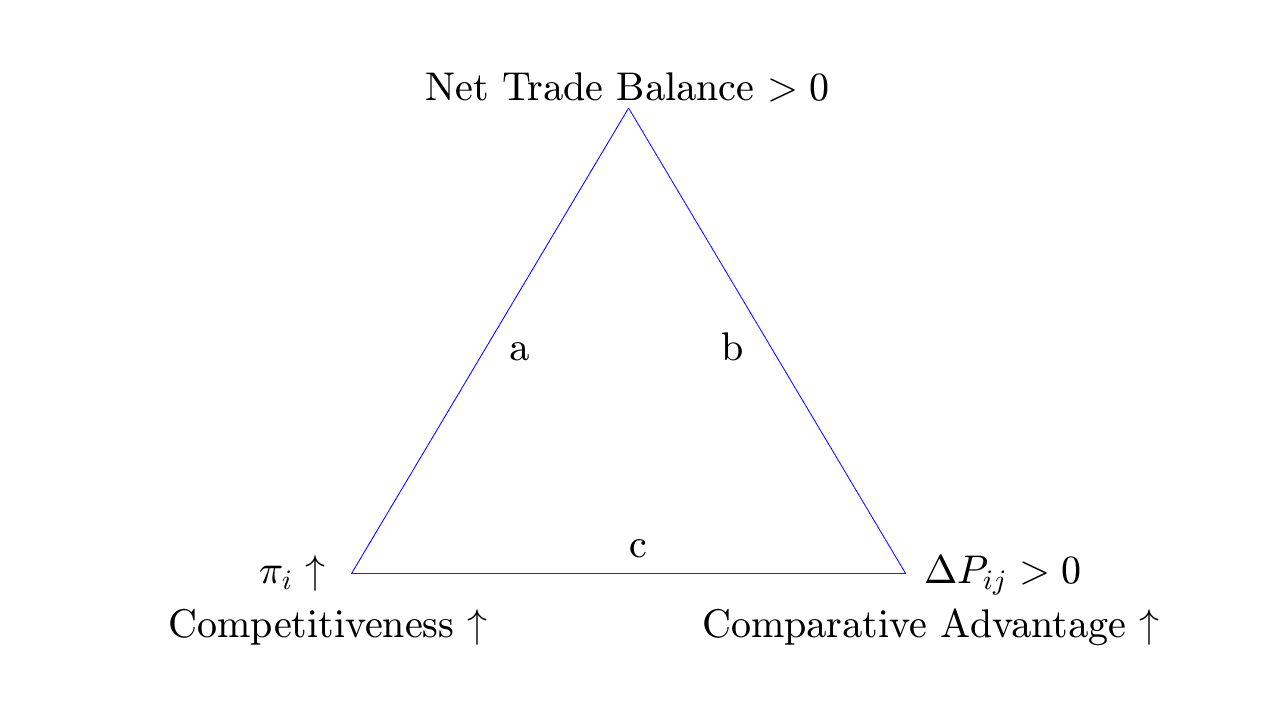}
\caption{impossibility trilemma of $\pi_i \uparrow$, trade surplus, and $\Delta P_{ij}>0$}\label{fig:Impossibility_Trilemma}
\end{figure}

Figure~\ref{fig:Impossibility_Trilemma} illustrates this trilemma: at most two of the three policy objectives can be achieved by country $i$. For example, if country $i$ adopts a position along edge $b$, it can overturn comparative advantage and secure a trade surplus, but only by sacrificing competitiveness --- unless it offsets this by absorbing competitiveness from other countries.
A historical example is the early 2000s, when bilateral exports between China and the USA grew as a share of their respective production. According to the trilemma, the USA had to either lose competitiveness, incur trade deficits, or both.

Country $i$'s interests in increasing competitiveness, maintaining a trade surplus, and achieving export growth are not perfectly aligned, and they do not completely contradict those of country $j$.
In Figure~\ref{fig:Impossibility_Trilemma}, country $i$'s optimal policy position lies somewhere within the triangle, while country $j$'s interests may lie elsewhere.
A mutually beneficial outcome occurs when $\Delta P_{ij}>0$ and $\Delta P_{ji}>0$, as both countries gain from economies of scale and declining marginal production costs.

According to the proof of Theorem~\ref{thm:impossibility_trilemma}, country $j$ could, under certain conditions,  achieve all three objectives simultaneously.
While this may seem unfair to country $i$, it is conditional on the Matthew effect, which inherently favors $\pi_i$. 
Thus, Theorem~\ref{thm:impossibility_trilemma} serves as a corrective measure to this bias.

\subsection{Nash Bargaining Solutions}

This subsection explores a compromise between trade balance and competitiveness.
In international trade, some countries prioritize competitiveness, while others focus on improving trade balance --- either by reducing deficits or increasing surpluses.
These preferences influence the choice of $\lambda_{ji}$, which can maintain either competitiveness $\pi_i$ or trade balance unchanged at $\lambda_{ji}=\pi_j/\pi_i$ or $\lambda_{ji}=g_j/g_i$, respectively.
If countries are indifferent between competitiveness and trade balance, a natural midpoint might seem like a solution:
$\lambda_{ji} = \left (g_j/g_i + \pi_j/\pi_i  \right) / 2$ and $\lambda_{ij} = \left (g_i/g_j + \pi_i/\pi_j\right) / 2$.
However, this midpoint does not satisfy the identity condition $\lambda_{ji} \lambda_{ij}=1$.

When countries $i$ and $j$ have opposing priorities --- one favoring competitiveness and the other trade balance --- the Nash bargaining solution (1950) provides a cooperative approach that balances $\pi_j/\pi_i$ and $g_j/g_i$ while ensuring $\lambda_{ji} \lambda_{ij}=1$.

Define:  
\begin{equation}\label{eq:def_p_q}
p \overset{\mathrm{def}}{=\joinrel=} \min \left \{ \frac{\pi_j}{\pi_i}, \frac{g_j}{g_i} \right \}
\quad \mathrm{and} \quad
q \overset{\mathrm{def}}{=\joinrel=} \max \left \{ \frac{\pi_j}{\pi_i}, \frac{g_j}{g_i}\right \}
\end{equation}
which are either $\pi_j/\pi_i$ or $g_j/g_i$. Clearly,
$$
\resizebox{.85\hsize}{!}{$
\frac{1}{p}= \frac{1}{\min \left \{ \frac{\pi_j}{\pi_i}, \frac{g_j}{g_i} \right \}} = \max \left \{ \frac{\pi_i}{\pi_j}, \frac{g_i}{g_j} \right \}
\quad
\mathrm{and}
\quad
\frac{1}{q}= \frac{1}{\max \left \{ \frac{\pi_j}{\pi_i}, \frac{g_j}{g_i}\right \}} = \min \left \{ \frac{\pi_i}{\pi_j}, \frac{g_i}{g_j}\right \}.
$}
$$

Without bargaining, according to Corollary~\ref{cor:zero_d_pi} and Theorem~\ref{thm:zero_trade_deficit}, $\lambda_{ji}$ can secure $p$ or $q$, and $\lambda_{ij}$ can secure $1/p$ or $1/q$. These represent their respective status quo points if no agreement is reached.

Two scenarios arise when countries $i$ and $j$ have conflicting preferences.
In the first scenario, country $j$ prefers $q$ over $p$, while $i$ prefers $1/p$ over $1/q$. 
However, $(q,1/p)$ is not a solution for $(\lambda_{ji},\lambda_{ij})$ since $q\times 1/p \neq 1$.
By Eq.~(\ref{eq:def_p_q}), $p$ and $1/q$ are their respective status quo points, obtained if one decides not to bargain with the other. 
In the second scenario, country $j$ prefers $p$ over $q$, while $i$ prefers $1/q$ over $1/p$. 
A preference over these alternatives defines a von Neumann-Morgenstern utility function for the bargaining problem.

Theorem~\ref{thm:Nash_bargaining} resolves these bargaining problems.
The solution satisfies the identity $\lambda_{ji} \lambda_{ij}=1$ by design.
It remains consistent across both scenarios and under any positive affine transformation of utility functions (a property of Nash bargaining).
By taking the square root of $\pi_j/\pi_i$, the solution also mitigates the Matthew effect in $\pi$,
bridging the gap between these $g_j/g_i$ and $\pi_j/\pi_i$.

\begin{theorem}\label{thm:Nash_bargaining}
When countries $i$ and $j$ have opposing preferences on $p$ and $q$, the Nash bargaining solution is: 
$$
\lambda_{ji}^* \overset{\mathrm{def}}{=\joinrel=} \sqrt{\frac{\pi_j g_j }{\pi_i g_i}} 
\hspace{.5cm}\mathrm{and}\hspace{.5cm}
\lambda_{ij}^* \overset{\mathrm{def}}{=\joinrel=} \sqrt{\frac{\pi_i g_i }{\pi_j g_j}}
$$ 
for $\lambda_{ji}$ and $\lambda_{ij}$, respectively.
\end{theorem}

\begin{figure}[ht]
\centering
\parbox{5cm}{
\centering
\includegraphics[height=4.5cm, width=6cm]{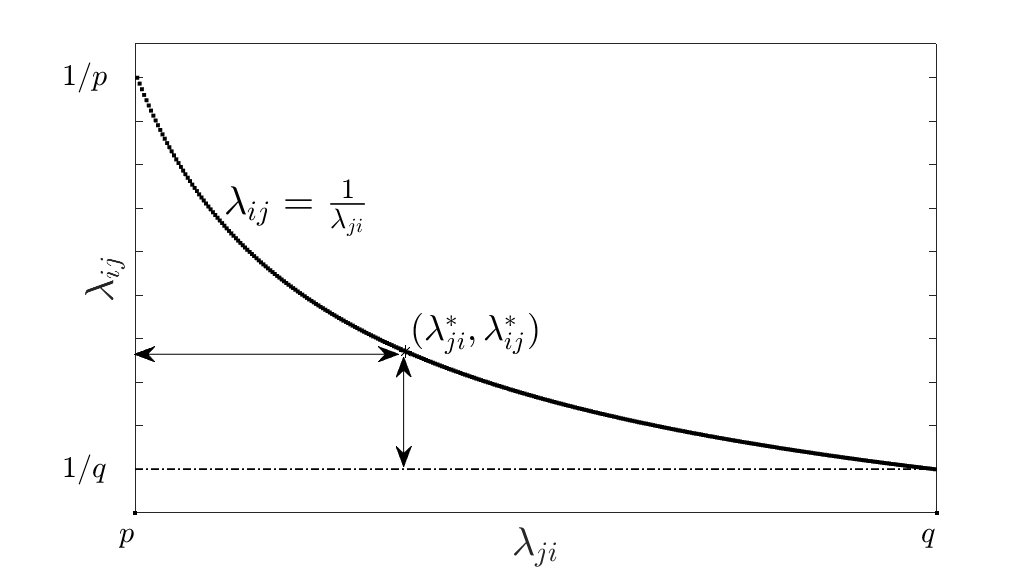}}
\quad
\begin{minipage}{6.5cm}
\centering
\includegraphics[height=4.5cm, width=6cm]{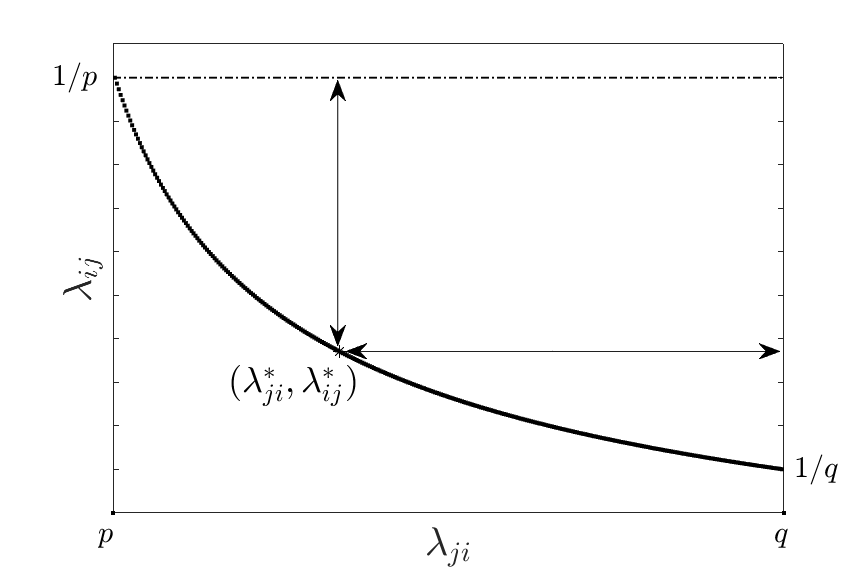}
\end{minipage}
\caption{Nash bargaining solution for $\lambda_{ji}^*$ when $i$ and $j$ have opposite preferences}\label{fig:Nash}
\end{figure}

Figure~\ref{fig:Nash} illustrates the bargaining solution.
In the first scenario (left plot), country $j$ moves from left to right along the curve $\lambda_{ij} = 1/\lambda_{ji}$, while country $i$ moves from bottom to top along the same curve.
They meet at the solution point $(\lambda^*_{ji}, \lambda^*_{ij})$, which maximizes the area of the rectangle formed by this point and the status quo point $(p, 1/q)$.
The second scenario (right plot) shows a similar process where where $j$ moves from $q$ to $p$ and $i$ moves from $1/p$ to $1/q$ along the curve.
Again, they meet at $(\lambda^*_{ji}, \lambda_{ij}^*)$, maximizing the area of the rectangle defined by this point and $(q, 1/p)$.

The solutions $\lambda_{ji}^*$ and $\lambda_{ij}^*$ depend on the choice of status quo points.
The effective bargaining set (or retaliation corridor) is $[p,q]$ for $\lambda_{ji}$ and $[1/q,1/p]$ for $\lambda_{ij}$, because country $j$ can secure $p$ (or $q$) and $i$ can secure $1/q$ (or $1/p$) in the first (or second) scenario. 
These corridors help prevent hostile overreactions and unintended consequences, such as third-party blowback.
However, they can be adjusted if both players symmetrically strengthen or weaken their status quo positions.

If we scale the two status quo points by a positive number $c\le \sqrt{q/p}$,
the effective bargaining set for  $\lambda_{ji}$ expands from $[p,q]$ to $[cp, q/c]$, 
and for $\lambda_{ij}$ from $[1/q,1/p]$ to $[c/q,1/(cp)]$.
Despite this scaling, the bargaining solution remains unchanged (Corollary~\ref{cr:Nash_bargaining}),
while the new effective bargaining sets approach $(0,\infty)$ as $c\to 0$.

\begin{corollary}\label{cr:Nash_bargaining}
For any positive $c \le \sqrt{q/p}$, replacing $p$ with $cp$ and $q$ with $q/c$ does not change the bargaining solution:  $\lambda_{ji} = \lambda_{ji}^*$ and $\lambda_{ij}=\lambda_{ij}^*$.
\end{corollary}

Additionally, as stated in Corollary~\ref{cr:Nash_confrontation}, the ratios ${\pi_j}/{\pi_i}$ and ${g_j}/{g_i}$ are also Nash bargaining solutions for $\lambda_{ji}$ when countries $i$ and $j$ directly compete for competitiveness and trade balance, respectively.
The solution $\lambda_{ji} = \lambda_{ji}^*$ or $\pi_j/\pi_i$ --- but not $g_j/g_i$ --- appears endogenous to the derivative $\frac{\mathrm{d} \pi_i}{\mathrm{d} P_{ji}}$, whose zero value provides a status quo point for the Nash solution. 
Nevertheless, the solution remains unchanged regardless of the derivative.

\begin{corollary}\label{cr:Nash_confrontation}
Both ${\pi_j}/{\pi_i}$ and ${g_j}/{g_i}$ are Nash bargaining solutions for $\lambda_{ji}$.
\end{corollary}

However, direct confrontation may be an inappropriate option when $g_i$ and $g_j$ differ significantly, according to Corollary~\ref{cr:homogeneous_preference}.
This aligns with China's strategy with the USA from 1980 to 2010, when Deng Xiaoping advocated ``keeping a low profile and taking no lead'' (e.g., Yan, 2014). 
Historically, an even distribution of economic, political, and military capabilities between rival countries tends to increase the likelihood of war, whereas peace is best preserved when there is an imbalance of national capabilities  (e.g., Organski, 1968).
Therefore, Corollary~\ref{cr:homogeneous_preference} suggests using $\lambda_{ji} = \lambda_{ji}^*$ as the rule of thumb, unless their sizes of economies are comparable --- in this case, $\frac{\pi_j}{\pi_i}$ and $\frac{g_j}{g_i}$ are also viable options for $\lambda_{ji}$.

\begin{corollary}\label{cr:homogeneous_preference}
During globalization era when $\Delta P_{ji}>0$, country $j$ should strategically avoid homogeneous preferences with a significantly larger country $i$.
\end{corollary}

\subsection{National Bargaining Power}
In deriving the solution $\lambda_{ji}^*$, both countries are assumed to have symmetric roles and identical prior bargaining power, 
consistent with the Nash bargaining axiom.
In practice, however, countries $i$ and $j$ typically exhibit asymmetric prior bargaining power.
For example, one could maximize 
\begin{equation}\label{eq:aNash}
\left(\lambda_{j i}-p\right)^{\zeta_{ij}}\left(\lambda_{i j}-1/q\right)^{1-\zeta_{ij}}
\quad \mathrm{or}\quad
\left(\lambda_{j i}-q\right)^{\zeta_{ij}}\left(\lambda_{i j}-1/p\right)^{1-\zeta_{ij}}
\end{equation} 
for some constant $\zeta_{ij} \in(0,1)$ (e.g., Anbarci and Sun, 2013).

In real-world bargaining, however, $\zeta_{ij}$ depends on numerous factors, including non-economic elements such as military strength, population, culture, legal systems, and geographic location. These factors, along with alternative optimal criteria, can significantly violate the symmetry axiom, resulting in bargaining outcomes that deviate from $\lambda_{ji}^*$ and $\lambda_{ij}^*$.
Without prior knowledge of $\zeta_{ij}$, there is no justification to favor one country over another, leading to the symmetric solution $\lambda^*_{ji}$, which may provide information about bargaining power. 

To address this, we introduce a posterior bargaining power measure that combines the symmetric prior with empirical data, building on Theorem~\ref{thm:Nash_bargaining}.
Extending the bilateral Nash solution to multilateral negotiations, we define:
$$
\beta_i \overset{\mathrm{def}}{=\joinrel=} \frac{\sqrt{\pi_i g_i}}{\sum\limits_{j \in \mathcal{N}}\sqrt{\pi_j g_j}}
$$
where $\beta_i$ represents country $i$'s national bargaining power in the trade system.
This measure is normalized so that the sum of all bargaining powers equals one, enabling comparisons over time.
It is particularly useful in multilateral context, such as
allocating vote rights in forming or reforming international alliances, as it avoids the need for bilateral negotiations.
Despite this, bilateral trade agreements remain more common than multilateral ones (e.g., Yilmazkuday and Yilmazkuday, 2014).

The set $\{\beta_i\}_{i=1}^n$ establishes a linear ordering of countries and preserves the transitivity property of the Nash solution $\lambda_{ij}^*$.
If $\beta_i > \beta_j$ and $\beta_j > \beta_k$, then $\beta_i > \beta_k$.
Specifically, country $i$'s bargaining power relative to $k$ equals:
$$
\lambda_{ik}^* = \lambda_{ij}^* / \lambda_{jk}^* = \beta_i / \beta_k, \quad \forall j\in \mathcal{N}.
$$

Two equally weighted and highly correlated factors determine $\beta_i$: production volume $g_i$ and competitiveness strength $\pi_i$.
Consequently, $\beta_i$ inherits some of the ``controversial'' properties of $\pi_i$.
From the production aspect, $\beta_i$ decreases when domestic firms rely heavily on imported inputs, reducing the country's leverage in negotiations.
From the consumption perspective, bargaining power rises when the country imports more goods and services, influencing foreign production.
However, the positive correlation between the breadth $g_i$ and depth $\pi_i$ of the economy introduces a risk of a reverse Matthew effect: a decline in $g_i$ can trigger a reduction in $\pi_i$, which in turn further depresses $g_i$.
As a result, $\beta_i$ may shrink more than the initial drop in $g_i$.

\section{Economic Globalization}\label{sect:globalization}

Globalization refers to the removal of barriers to the flow of financial products, goods, technology, information, and jobs across national borders and cultures. 
In economic terms, it describes the growing interdependence among countries, driven by trade in goods and services.
A notable example is China's accession to the WTO in 2001, while the UK's Brexit in 2020 represented a clear rejection of globalization.
Conversely, protectionist policies can raise tariffs on imports, impose quotas, and enforce stricter regulations. 
For instance, in 2018, the Trump administration imposed punitive tariffs on all steel and aluminum imports (Washington Post, 2018b).

Globalization creates both winners and losers within nations, but it does not automatically compensate those who lose, often fueling income inequality and political divisions.
Therefore, a more inclusive approach is needed --- one that not only maximizes the gains of winners but also mitigates negative impacts on others. 
The benefits of globalization are also unevenly distributed across countries.

This section explores whether a nation should retreat from global economic integration without directly targeting any specific country. 
The goal is to enhance national competitiveness rather than simply maximizing the aggregate utility of trade firms. 
Our analysis focuses on actions that influence the global economy as a whole, while disregarding bilateral trade balances.

\subsection{Optimal Level of Globalization}
When country $i$ takes an action to adjust its level of globalization, it triggers changes across the entire matrix $P$.
The country's first move affects the diagonal element $P_{ii}$ with a small change $\Delta P_{ii}$.
If globalization increases, $P_{ii}$ decreases (i.e., $\Delta P_{ii}<0$); if protectionism rises, $P_{ii}$ increases (i.e., $\Delta P_{ii}>0$).

Since the action does not target specific countries, we assume that
whenever column $i$ of $P$ changes, the other columns adjust proportionally to maintain the unit sum of each row.
For example, a change $\Delta P_{ji}$ induces proportional changes $\Delta P_{ji} \sigma_{ji}$ within row $j$ where:
$$
\sigma_{ji} \overset{\mathrm{def}}{=\joinrel=} \frac{1}{1-P_{ji}}\left(-P_{j1},\cdots, -P_{j,i-1}, 1-P_{ji}, -P_{j,i+1},\cdots, -P_{jn}\right).
$$
Consequently, a change $\Delta P_{ii}$ in $P$ produces three levels of adjustments $\Delta P$ in $P$. 
First, the $i$th row in $\Delta P$ is $\Delta P_{ii} \sigma_{ii}$.
Secondly, the $i$th column in $\Delta P$ is $\left[\Delta P_{ii} \sigma_{ii} \odot \Lambda_i\right]^\top$, due to reprisal reactions, where $\odot$ denotes the Hadamard (element-wise) product and $\Lambda_i$ is the $i$th row of $\Lambda$.
Lastly, the change in column $i$ propagates to all other columns.

We define:  
$$
M \overset{\mathrm{def}}{=\joinrel=} \mathrm{diag} \left( \sigma_{ii} \odot \Lambda_i \right)
\left[\begin{array}{c}\sigma_{1i} \\ \vdots \\ \sigma_{ni} \end{array} \right]
$$
where $\mathrm{diag}(\sigma_{ii} \odot \Lambda_i)$ is the diagonal matrix with $\sigma_{ii} \odot \Lambda_i$ on its diagonal.
Thus, the $j$th row of $M$ is $\sigma_{ji}$, scaled by the $j$th element of $\sigma_{ii} \odot \Lambda_i$.
Therefore: 
$$
\Delta P = \Delta P_{ii} M, \quad
\mathrm{and} \ \mathrm{as}\ \Delta P_{ii}\to 0,\quad
\frac{\mathrm{d} P}{\mathrm{d} P_{ii}} = M.
$$ 
Finally, let $M_i$ denote $M$ with its $i$th column dropped.
Theorem~\ref{thm:globalization} provides the derivative of $\pi$ with respect to $P_{ii}$.

\begin{theorem}\label{thm:globalization}
Under the above setting and notations, we have:
\begin{equation} \label{eq:pi_i_2_P_ii}
\frac{\mathrm{d} \pi_i}{\mathrm{d} P_{ii}} 
= 
-\frac{\vec{1}^\top_{n-1} (I_{n-1} -Z_i)^{-1} (\pi M_i)^\top}{1+\vec{1}_{n-1}^\top (I_{n-1}-Z_i)^{-1} \alpha_i}
\end{equation}
and
\begin{equation} \label{eq:pi_minus_i_2_P_ii}
\frac{\mathrm{d} \pi_{-i}}{\mathrm{d} P_{ii}}
=
\left(I_{n-1}-Z_i \right)^{-1}
\left [ 
\left(\pi M_i\right)^\top - \frac{\vec{1}^\top_{n-1} (I_{n-1} -Z_i)^{-1} (\pi M_i)^\top}{1+\vec{1}_{n-1}^\top (I_{n-1}-Z_i)^{-1} \alpha_i} \alpha_i
\right ].
\end{equation}
\end{theorem}

The decision rule on globalization could be as follows: country $i$ should pursue further globalization or protectionism if $\frac{\mathrm{d} \pi_i}{\mathrm{d} P_{ii}}$ is significantly negative or positive, respectively.
If this derivative is insignificantly different from zero, country $i$ may instead seek trade war targets or trade partnerships as outlined in Sections~\ref{sect:trade_war} and \ref{sect:bargaining}. In addition,
a positive (negative) sign of $\frac{\mathrm{d} \pi_i}{\mathrm{d} P_{ii}}$ indicates the country is over-globalized (under-globalized, respectively).

The optimal level of globalization for country $i$ maximizes $\pi_i$ by adjusting $P_{ii}$, thereby rebalancing $P$ by $\Lambda$.
The first-order condition $\frac{\mathrm{d} \pi_i}{\mathrm{d} P_{ii}}=0$ is necessary at the optimum level because  $P_{ii}$ cannot be $1$ or $0$.
However, $\frac{\mathrm{d} \pi_j}{\mathrm{d} P_{ii}}$ may remain positive or negative at this optimal level, allowing country $j$ to exploit this by adjusting $P_{ij}$, thereby influencing $P_{ii}$.
To deter simultaneous strategic moves by other countries, we impose $\frac{\mathrm{d} \pi}{\mathrm{d} P_{ii}} = \vec{0}_n^\top$.
This condition holds if and only if $\pi M_i=\vec{0}_{n-1}$, as stated in Corollary~\ref{cr:zero_pi_pii}.

\begin{corollary}\label{cr:zero_pi_pii}
$\frac{\mathrm{d} \pi}{\mathrm{d} P_{ii}}=\vec{0}^\top_n$ if and only if $\pi M_i = \vec{0}^\top_{n-1}$.
\end{corollary}

Given $\Lambda$, Eqs.~(\ref{eq:pi_i_2_P_ii}) and (\ref{eq:pi_minus_i_2_P_ii}) for all $i \in \mathcal{N}$ define a dynamical system that describes the instantaneous change of $\pi$ with respect to the diagonal elements of $P$.
The system has $n$ independent variables since the proportions among off-diagonal elements remain unchanged.
Due to its complexity, numerical algorithms can approximate a solution to $\frac{\mathrm{d} \pi}{\mathrm{d} P_{ii}}=\vec{0}_n^\top$ for all $i \in \mathcal{N}$ through successive small adjustments on the diagonal elements.
For example, over-globalized countries may increase $P_{ii}$ by 1\% at each step, while under-globalized countries reduce $P_{ii}$ by 1\% simultaneously. This iterative process continues until both the matrix $P$ and the vector $\pi$ converge. 
Unlike the convergences discussed in Lau et al. (2022), the diagonal elements do not necessarily converges to the same value.

Alternatively, we examine the properties of the solution to the first-order conditions.
If such a solution exists, it satisfies a new counterbalance equilibrium given in Eq.~(\ref{eq:steady_state_pi_lambda}), as stated in Corollary~\ref{cr:zero_pi_pii_all}. 
Eq.~(\ref{eq:steady_state_pi_lambda}) imposes $n-1$ identities to the solution, requiring only one additional condition to uniquely identify it.
Moreover, $P$ and $\pi$ already satisfy the new counterbalance if $\Lambda$ exhibits certain desirable properties, such as $\lambda_{ij}=\pi_i/\pi_j$ or $P_{ji}/P_{ij}$, which correspond to balanced bilateral authority or trade volumes according to Eq.~(\ref{eq:pi_i_pi_j_breakeven}) and Theorem~\ref{thm:zero_trade_deficit}, respectively.
Notably, Eq.~(\ref{eq:steady_state_pi_lambda}) reduces to Eq.~(\ref{eq:equilibrium}) when $\lambda_{ji}=P_{ij}/P_{ji}$, indicating no globalization or deglobalization is likely occur in the absence of trade imbalance.

\begin{corollary}\label{cr:zero_pi_pii_all}
If $\frac{\mathrm{d} \pi}{\mathrm{d} P_{ii}}=\vec{0}^\top_{n}$ for all $i\in \mathcal{N}$, then $\pi$ satisfies:
\begin{equation}\label{eq:steady_state_pi_lambda}
\pi \left[ \Lambda \odot P \right]^\top = \pi.
\end{equation}
Eq.~(\ref{eq:steady_state_pi_lambda}) holds if all $\lambda_{ij}=\pi_i/\pi_j$ or all $\lambda_{ij}=P_{ji}/P_{ij}$.
\end{corollary}

At the stable solution, any small perturbation to the diagonal of $P$ has a insignificant effect on $\pi$, discouraging countries from making minor adjustment.
However, this solution is optimal for some countries only and does not necessarily maximize competitiveness for all.
Additionally, the derivatives in Eqs.~(\ref{eq:pi_i_2_P_ii}) and (\ref{eq:pi_minus_i_2_P_ii}) rely on the exogeneity assumption of $\Lambda$ and proportional rebalancing in $M$.
Therefore, pursuing a comprehensive solution to the first-order conditions may have limited practical use.
Nevertheless, Theorem~\ref{thm:globalization} provides each country with guidance on the appropriate direction to move.

\subsection{Mixed Globalization--Protectionism Strategy}
In addition to globalization and trade wars, country $i$ can also form or join preferential trade agreements (PTAs) or regional trade agreements (RTAs) with other economies.
Ideally, within a PTA or RTA, the ratios $\lambda_j/\lambda_i$ and $g_j/g_i$ should remain balanced to facilitate consensus and prevent significant internal trade disputes.
Geographic and cultural proximity often encourage the formation of such agreements.
Moreover, Eq.~(\ref{eq:pi_minus_i_2_P_ii}) measures the external impact of country $i$'s protectionist policies on the rest of the world, aiding in the selection of suitable partners for PTAs or RTAs.
While some countries benefit from these, others may not.

Theorem~\ref{thm:globalization} suggests a mixed globalization--protectionism strategy, guiding country $i$ in deciding which PTAs or RTAs to join.
To implement this, we divide the countries in $\mathcal{N}\setminus \{ i \}$ into two groups based on the sign of their components in the vector $\pi M_i$:
\begin{itemize}
\item $\mathcal{N}_i^+$: countries with positive components;
\item $\mathcal{N}_i^-$: countries with negative components.
\end{itemize}

Since $(I_{n-1}-Z_i)^{-1}$ is a non-negative matrix and $\alpha_i$ is a non-negative vector, we rewrite Eq.~(\ref{eq:pi_i_2_P_ii}) as: $\frac{\mathrm{d} \pi_i}{\mathrm{d} P_{ii}} = - (\pi M_i) \omega$
for some non-negative vector $\omega$. 
Thus, when $j\in \mathcal{N}_i^+$ or $j\in \mathcal{N}_i^-$, its contribution to $\frac{\mathrm{d} \pi_i}{\mathrm{d} P_{ii}}$ is negative or positive, respectively, partially offsetting each other.
Given $\Delta P_{ii}$, the approximate change in $\pi_i$ is:
\begin{equation}\label{eq:approximate_global}
\Delta \pi_i \approx -\sum_{j\in \mathcal{N}_i^+}  \left(\pi M_i\right)_{_j} \omega_j \Delta P_{ii} - \sum_{j\in \mathcal{N}_i^-} \left(\pi M_i\right)_{_j} \omega_j \Delta P_{ii}.
\end{equation}

Under the mixed strategy, country $i$ increases protectionism toward countries in $\mathcal{N}_i^-$ and deepens globalization with countries in $\mathcal{N}_i^+$, thereby avoiding the partial offsetting effect.
This approach outperforms a purely globalization or purely protectionism strategy because the right-hand side of Eq.~(\ref{eq:approximate_global}) is less than or equal to:
$$
-\sum\limits_{j\in \mathcal{N}_i^+}  (\pi M_i)_{_j} \omega_j (-|\Delta P_{ii}|) - \sum\limits_{j\in \mathcal{N}_i^-}  (\pi M_i)_{_j} \omega_j |\Delta P_{ii}|.
$$
Therefore, if most members of a PTA or RTA belong to $\mathcal{N}_i^+$, then country $i$ should consider joining that  agreement. However, note that $j\in \mathcal{N}_i^+$ does not necessarily imply $i\in \mathcal{N}_j^+$, which may require additional negotiations if country $j$ is already part of the PTA or RTA.

\section{Empirical Study}\label{sect:empirical}

In this empirical study, we apply our theoretical framework to real-world trade data spanning from 2000 to 2019.
This analysis offers quantitative evidence that supports our earlier arguments, making the theoretical insights more concrete and relevant for policymakers.
The empirical findings reveals evolving trends in national competitiveness and bargaining power, illustrating how various countries have gained or lost ground over time. 
These insights provide a historical lens through which to examine the impacts of globalization and trade conflicts.

Between 2000 and 2019, the global economy experienced profound transformations and challenges, with varying outcomes across regions. 
This period saw a marked increase in global trade volumes, fueled by rapid technological advancements, the expansion of international trade agreements, and the emergence of new economic powers. 
A pivotal moment was China's accession to the WTO in 2001, which triggered a surge in global trade activity.

The rise of emerging markets --- particularly China and India --- was instrumental in driving global economic growth.
However, it also contributed to escalating trade tensions, most notably between China and the USA. 
The 2008 financial crisis and the subsequent European debt crisis were major inflection points, causing widespread economic downturns and prompting significant policy interventions by governments and central banks worldwide. 

As globalization intensified and trade volumes expanded, nations increasingly demanded the world's reserve currency, primarily through trade surpluses. 
This demand led to appreciation of the reserve currency, which in turn reduced the currency issuer's export competitiveness and widened trade deficits.
Thus, reversing globalization trends could be a viable option for the issuer.
For the USA, reshoring manufacturing sectors could reduce imports and deficits.
For China, sustained economic growth is essential to narrow the competitiveness gap with the USA, in line with the Matthew effect in $\pi$.

This empirical analysis calculates $\pi$ and its derivatives with respect to $P_{ii}$ and $P_{ji}$.
Bilateral trade data are sourced from the United Nations ComTrade database (2021),
while import and export data for non-consumer goods come from the World Bank's WITS Database (2021).
For comparison, we focus on two benchmark years: 2000, prior to China's accession to the WTO in 2001, and 2017, just before the onset of the China-U.S. trade war in 2018. 

Although the calculations include all economies in the dataset, we present results for China, Russia, and the G7 countries (Canada, France, Germany, Italy, Japan, the UK, and the USA), identified by
their ISO-3 codes: CHN, RUS, CAN, FRA, DEU, ITA, JPN, GBR, and USA.
The computations use the default coefficients, $\lambda_{ji}=\lambda_{ji}^*$, and
the results are summarized in the following tables and figures.

\begin{figure}[ht]
\centering
\includegraphics[height=6.8cm, width=13cm]{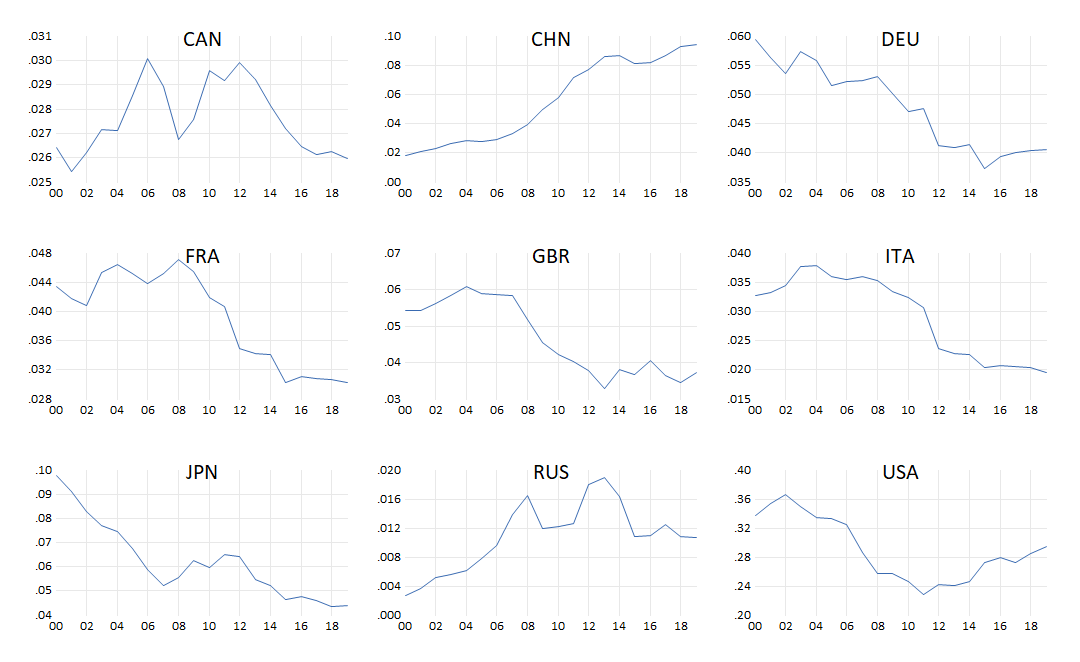}
\caption{Competitiveness $\pi_i$ for China, Russia, and G7 countries (2000-2019)}\label{fig:pi}
\end{figure}

Figure~\ref{fig:pi} illustrates competitiveness trends from 2000 to 2019.
During this period, China and Russia recorded significant gains, while Canada remained relatively stable, and the other six countries experienced declines.
Notably, the USA's competitiveness rose steadily until 2003, then dropped sharply through 2011, followed by a gradual recovery --- resulting in an overall decline of 12.7\%.
China's competitiveness surged during the two major crises but otherwise plateaued.
Canada and Russia demonstrated resilience, supported by rising commodity prices.
The four Western European countries were severely affected by the debt crisis, with competitiveness falling until 2015.
Germany's decline began in 2000, while the other three started in 2008.
Japan's competitiveness nearly halved in the first eight years, then stabilized between 2007 and 2014.
From 2000 to 2007, Japan faced a reverse Matthew effect as China and South Korea challenged its automobile and electric industries.

\begin{figure}[ht]
\centering
\parbox{5cm}{
\centering
\includegraphics[height=3cm, width=4.5cm]{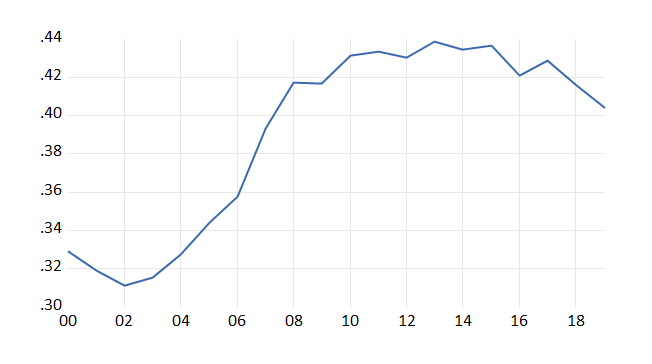}
\caption{Sum of $\pi_i$ for all other countries (2000-2019)}\label{fig:pi_others}}
\hspace{.8cm}
\begin{minipage}{5cm}
\centering
\includegraphics[height=3cm, width=4.5cm]{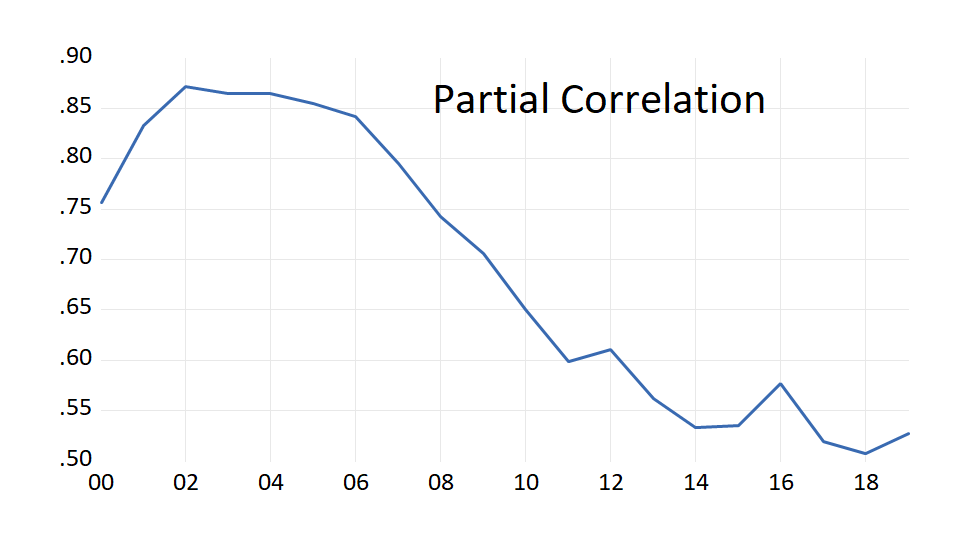} 
\caption{Correlation between $\pi_i$ and $g^2_i$ after controlling for $g_i$ (2000-2019)}\label{fig:correlation}
\end{minipage}
\end{figure}

Beyond these nine countries, other economies collectively improved competitiveness by 22.7\%, as shown in Figure~\ref{fig:pi_others}.
Many countries in Southern Asian, Southeast Asian, and the Middle East --- including India, Indonesia, Iraq, the United Arab Emirates, and Vietnam --- tripled their competitiveness.

The Matthew effect is clearly evident in the data. 
For example, in 2018, the production ratio between the USA and China was $1.344$, but their competitiveness ratio reached $3.070$. This disparity underscores the cumulative advantage phenomenon.
Further evidence in presented in Figure~\ref{fig:correlation}, which illustrates a strong positive correlation between $\pi_i$ and $g_i^2$, after controlling for $g_i$.
However, the partial correlation declined from $.757$ to $.527$ over the study period, coinciding with the transition of the trade system from a less multipolar to a more multipolar structure.

\renewcommand{\arraystretch}{1.4}
\begin{table}[ht]
\caption{Country $j$'s direct contribution to $\pi_i$ for 2017 and 2019 (in percentage)$^*$}
\label{tbl:contribution_distribution}
\centering
\resizebox{.95\hsize}{!}{
\begin{tabular}{r||r |r |r |r |r |r |r |r |r||c} \hline \hline
\rotatebox{25}{\Large{${}_{i}\mkern-6mu\setminus\mkern-6mu{}^{j}$}}&CAN	&CHN	&DEU	&FRA	&GBR	&ITA	&JPN	&RUS	&USA&$\pi_i$\\ \hline		
CAN	&$\frac{76.647}{77.390}$	&$\frac{ 1.2635}{ 1.2732}$	&$\frac{ 0.5390}{ 0.5179}$	&$\frac{ 0.3222}{ 0.2781}$	&$\frac{ 0.5922}{ 0.5728}$	&$\frac{ 0.2877}{ 0.2607}$	&$\frac{ 0.4329}{ 0.4768}$	&$\frac{ 0.0410}{ 0.0355}$	&$\frac{14.492}{13.916}$&$\frac{.025965}{.026125}$\\		
CHN	&$\frac{ 0.3952}{ 0.3439}$	&$\frac{85.584}{84.126}$	&$\frac{ 0.8233}{ 0.8565}$	&$\frac{ 0.3967}{ 0.3931}$	&$\frac{ 0.3423}{ 0.3479}$	&$\frac{ 0.1878}{ 0.2052}$	&$\frac{ 1.3131}{ 1.4868}$	&$\frac{ 0.3110}{ 0.3095}$	&$\frac{ 1.5849}{ 2.1950}$&$\frac{.093905}{.086596}$\\		
DEU	&$\frac{ 0.2088}{ 0.1682}$	&$\frac{ 1.2344}{ 1.2456}$	&$\frac{67.501}{67.410}$	&$\frac{ 2.2469}{ 2.3993}$	&$\frac{ 1.8879}{ 2.0095}$	&$\frac{ 1.4928}{ 1.5634}$	&$\frac{ 0.4265}{ 0.3507}$	&$\frac{ 0.3737}{ 0.5379}$	&$\frac{ 3.2504}{ 3.0282}$&$\frac{.040521}{.040105}$\\
FRA	&$\frac{ 0.2345}{ 0.2171}$	&$\frac{ 0.7032}{ 0.6617}$	&$\frac{ 3.5649}{ 3.5687}$	&$\frac{74.940}{75.190}$	&$\frac{ 1.9053}{ 1.8352}$	&$\frac{ 1.6149}{ 1.5794}$	&$\frac{ 0.2216}{ 0.2111}$	&$\frac{ 0.1656}{ 0.2247}$	&$\frac{ 2.4124}{ 2.3005}$&$\frac{.030196}{.030772}$\\		
GBR	&$\frac{ 0.6234}{ 0.6356}$	&$\frac{ 0.9550}{ 0.9428}$	&$\frac{ 2.2098}{ 2.4071}$	&$\frac{ 1.5773}{ 1.6503}$	&$\frac{75.873}{75.862}$	&$\frac{ 0.6921}{ 0.7259}$	&$\frac{ 0.3716}{ 0.4077}$	&$\frac{ 0.1986}{ 0.1736}$	&$\frac{ 4.0273}{ 3.5605}$&$\frac{.037385}{.036390}$\\		
ITA	&$\frac{ 0.1418}{ 0.1419}$	&$\frac{ 1.0159}{ 0.9305}$	&$\frac{ 3.2895}{ 3.1651}$	&$\frac{ 2.3107}{ 2.3567}$	&$\frac{ 1.3228}{ 1.2695}$	&$\frac{74.945}{75.662}$	&$\frac{ 0.1958}{ 0.2123}$	&$\frac{ 0.4274}{ 0.4525}$	&$\frac{ 1.7055}{ 1.5402}$&$\frac{.019532}{.020507}$\\		
JPN	&$\frac{ 0.3297}{ 0.3118}$	&$\frac{ 2.0275}{ 1.9912}$	&$\frac{ 0.5242}{ 0.4869}$	&$\frac{ 0.3184}{ 0.3041}$	&$\frac{ 0.4209}{ 0.3614}$	&$\frac{ 0.2218}{ 0.2139}$	&$\frac{87.560}{87.198}$	&$\frac{ 0.1581}{ 0.1924}$	&$\frac{ 2.2315}{ 1.9753}$&$\frac{.043971}{.045859}$\\		
RUS	&$\frac{ 0.1182}{ 0.1099}$	&$\frac{ 2.6727}{ 2.2351}$	&$\frac{ 1.7330}{ 1.7824}$	&$\frac{ 0.9775}{ 1.0709}$	&$\frac{ 0.7020}{ 0.7266}$	&$\frac{ 0.8945}{ 0.8129}$	&$\frac{ 0.6044}{ 0.5044}$	&$\frac{80.225}{79.511}$	&$\frac{ 1.5211}{ 1.6313}$&$\frac{.010657}{.012509}$\\		
USA	&$\frac{ 1.3878}{ 1.4971}$	&$\frac{ 0.8107}{ 1.0322}$	&$\frac{ 0.4527}{ 0.4749}$	&$\frac{ 0.2580}{ 0.2740}$	&$\frac{ 0.5935}{ 0.5643}$	&$\frac{ 0.1733}{ 0.1802}$	&$\frac{ 0.3514}{ 0.3897}$	&$\frac{ 0.0368}{ 0.0464}$	&$\frac{92.142}{91.369}$&$\frac{.293880}{.272142}$\\
\hline\hline
\multicolumn{11}{l}{\small * The numerators correspond to 2019 values, and the denominators to 2017. } \\
\end{tabular}
}
\end{table}

A direct implication of Eq.~(\ref{eq:power_in}) is that country $j$'s contribution to country $i$'s competitiveness can be expressed as a percentage: $\pi_j P_{ji} / \pi_i \times 100\%$.
Table~\ref{tbl:contribution_distribution} reports these contributions for 2017 and 2019.
The diagonal values indicate that the USA was the most self-sufficient in sustaining its competitiveness, with a slight increase from 2017 to 2019.
The ninth column further shows that, in 2019, the USA was the largest contributor to Canada, China, Germany, Japan, and the UK.
In the same year, China was the primary contributor to Russia's competitiveness and the second-largest contributor to  Canada, Japan, and the USA.
France, Germany, and Italy exhibited strong mutual reliance, while their average dependence on Russia was 8.7 times higher than the USA's dependence on Russia.

Bilateral break-even points in Eq.~(\ref{eq:pi_i_pi_j_breakeven}), in general, do not exist. 
For example, in 2019, the USA contributed $.14492*.025965=.003763$ to Canada, whereas Canada contributed $.013878*.293880=.004078$ to the USA.
Comparing the periods before and after the onset of the China-USA trade war, the USA's contributions to China decreased by $21.7 \%$, calculated as $1-(1.5849*.093905)/(2.1950*.086596)$,
while China's contribution to the USA fell by $15.2\%$, calculated as $1 - (.8107*.29388)/(1.0322*.272142)$.

\begin{figure}[ht]
\centering
\includegraphics[height=6.8cm, width=13cm]{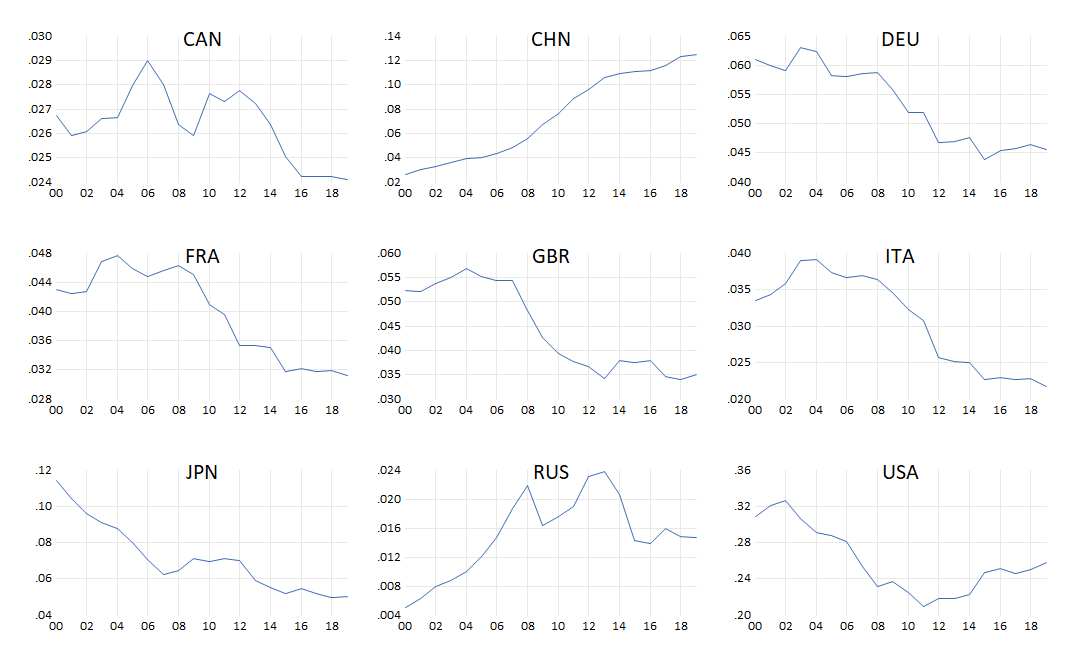}
\caption{Bargaining power $\beta_i$ (2000-2019)}\label{fig:bargain}
\end{figure}

Figure~\ref{fig:bargain} illustrates national bargaining power from 2000 to 2019,
showing patterns similar to Figure~\ref{fig:pi} but smoother.
China's bargaining power rose steadily due to rapid production growth, even during troughs in its $\pi_i$ in 2005 and 2015. 
Notably, China's bargaining power consistently exceeded its competitiveness, whereas the USA's bargaining power was consistently lower than its competitiveness --- evidence of $\beta$'s role in mitigating the Matthew effect in $\pi$.
Russia's bargaining power declined sharply after the Russo-Georgian War in 2008 and the annexation of Crimea in 2014.

\begin{figure}[ht]
\centering
\includegraphics[height=5cm, width=13cm]{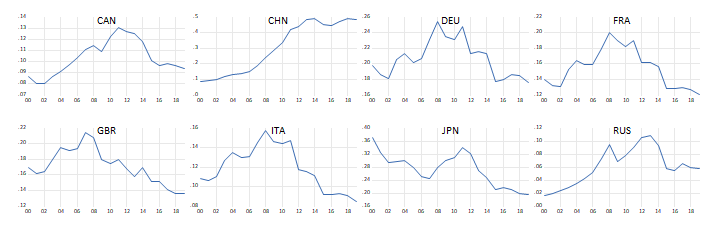} 
\caption{Bargaining power \textit{versus} the USA (2000-2019)}\label{fig:bargain_vs_usa}
\end{figure}

Figure~\ref{fig:bargain_vs_usa} compares the USA's bargaining power against the other countries.
Western European countries saw a slight increase after the Eurozone's formation in 1999, followed by a modest decline post-2008, resulting in an average 15.7\% loss of bargaining power relative to the USA.
Conversely, China and Russia significantly strengthened their relative power during the first 15 years, then stabilized.
Japan, however, experienced a substantial decline, losing about 50\% of its relative bargaining power.
Canada's bargaining strength spiked during crises but quickly reverted to its original level, reflecting its strong alignment with the USA.

\renewcommand{\arraystretch}{1.4}
\begin{table}[ht]
\caption{$1,000\times \frac{\mathrm{d} \log \pi_i}{\mathrm{d} \log P_{ji}}$ for years 2000 and 2017$^*$}
\label{tbl:d_pi_pji}
\centering
\resizebox{.97\hsize}{!}{
\begin{tabular}{r||r |r |r |r |r |r |r |r |r} \hline \hline
\rotatebox{25}{\Large{${}_{i}\mkern-6mu\setminus\mkern-6mu{}^{j}$}}&CAN	&CHN	&DEU	&FRA	&GBR	&ITA	&JPN	&RUS	&USA\\ \hline		
CAN	&&$\frac{  -25.78}{   -5.78}$	&$\frac{   -5.49}{   -0.37}$	&$\frac{   -1.47}{    0.29}$	&$\frac{   -0.72}{    2.07}$	&$\frac{   -2.35}{   -0.16}$	&$\frac{   -4.87}{   -5.35}$	&$\frac{   -0.62}{   -0.53}$	&$\frac{   15.75}{   63.78}$\\		
CHN	&$\frac{    6.58}{    6.90}$	&&$\frac{    7.39}{   16.62}$	&$\frac{    5.36}{    7.76}$	&$\frac{    6.05}{    8.46}$	&$\frac{    2.14}{    5.25}$	&$\frac{   11.56}{   31.43}$	&$\frac{    0.89}{   -2.48}$	&$\frac{   45.19}{   65.48}$\\		
DEU	&$\frac{    1.18}{    0.14}$	&$\frac{   -7.73}{   -6.05}$	&&$\frac{    6.49}{    3.43}$	&$\frac{   10.07}{    5.30}$	&$\frac{    1.29}{    0.18}$	&$\frac{    0.10}{   -5.13}$	&$\frac{   -1.86}{   -4.59}$	&$\frac{   23.54}{   15.06}$\\		
FRA	&$\frac{    0.98}{   -0.19}$	&$\frac{   -8.51}{   -5.16}$	&$\frac{  -13.75}{   -6.39}$	&&$\frac{    5.54}{    2.67}$	&$\frac{   -4.20}{   -2.57}$	&$\frac{   -0.90}{   -3.67}$	&$\frac{   -2.08}{   -3.10}$	&$\frac{   12.90}{    9.72}$\\		
GBR	&$\frac{    0.73}{   -0.96}$	&$\frac{  -16.47}{   -7.66}$	&$\frac{  -18.11}{   -7.01}$	&$\frac{   -5.31}{   -2.28}$	&&$\frac{   -4.64}{   -2.58}$	&$\frac{   -3.39}{   -7.65}$	&$\frac{   -2.38}{   -1.99}$	&$\frac{    8.25}{   10.81}$\\		
ITA	&$\frac{    1.01}{    0.14}$	&$\frac{   -8.29}{   -5.79}$	&$\frac{   -3.66}{   -0.48}$	&$\frac{    6.12}{    3.66}$	&$\frac{    7.17}{    4.41}$	&&$\frac{   -0.21}{   -2.97}$	&$\frac{   -2.65}{   -6.84}$	&$\frac{   12.51}{   11.97}$\\		
JPN	&$\frac{    4.31}{    3.52}$	&$\frac{  -26.28}{  -14.31}$	&$\frac{   -0.28}{    4.76}$	&$\frac{    2.01}{    3.72}$	&$\frac{    4.33}{    6.66}$	&$\frac{    0.35}{    2.27}$	&&$\frac{   -1.77}{   -2.08}$	&$\frac{   30.88}{   69.74}$\\		
RUS	&$\frac{    1.52}{    4.49}$	&$\frac{   -5.95}{    3.48}$	&$\frac{    8.91}{   56.17}$	&$\frac{    9.85}{   24.14}$	&$\frac{    9.05}{   26.41}$	&$\frac{    5.10}{   25.58}$	&$\frac{    2.78}{    6.37}$	&&$\frac{   24.25}{   56.37}$\\		
USA	&$\frac{   -1.78}{   -5.56}$	&$\frac{  -48.89}{  -17.13}$	&$\frac{  -12.58}{   -6.53}$	&$\frac{   -4.11}{   -2.91}$	&$\frac{   -3.01}{   -2.77}$	&$\frac{   -4.26}{   -2.95}$	&$\frac{  -10.14}{  -23.53}$	&$\frac{   -2.18}{   -1.84}$	&\\		
\hline \hline	
\multicolumn{10}{l}{\small * The numerators are for 2017 and the denominators for 2000. } \\
\end{tabular}
}
\end{table}

Table~\ref{tbl:d_pi_pji} reports the derivatives of $\log \pi_i$ with respect to $\log P_{ji}$ for the years 2000 and 2017, calculated using Eqs.~(\ref{eq:pi_i_2_P_ji}) and (\ref{eq:percentchange}).
For example, a 1\% reduction in U.S. imports from China in 2017 would have increased U.S. competitiveness by .04889\%.
Although the USA may have had reasons to initiate trade disputes with all eight other countries, its primary targets were China in 2017 and Japan in 2000, as indicated by the numerators and denominators in the last row.
The last column shows that China and Japan would have been the two largest beneficiaries had they strengthened collaboration with the U.S. in 2017 and 2000, respectively.
However, comparing the numerators and denominators across years reveals a significant decline in potential benefits from 2000 to 2017.
Moreover, any of these eight countries initiating trade friction against the USA would incur losses.
A trade war with Russia, in particular, would be almost fruitless due to the small values in the eighth column.
Finally, the numbers symmetric to the diagonal exhibit opposite signs, confirming Corollary~\ref{cor:opposite_sign}.

\renewcommand{\arraystretch}{1.4}
\begin{table}[ht]
\caption{$1,000\times \frac{\mathrm{d} \log \pi_j}{\mathrm{d} \log P_{ii}}$ for years 2000 and 2017$^*$}
\label{tbl:d_pi_Pii}
\centering
\resizebox{.97\hsize}{!}{
\begin{tabular}{r|| r | r | r | r | r | r | r | r |r} \hline \hline
\rotatebox{25}{\Large{${}_{i}\mkern-6mu\setminus\mkern-6mu{}^{j}$}}	&CAN	&CHN	&DEU	&FRA	&GBR	&ITA	&JPN	&RUS	&USA\\ \hline		
CAN	&$\frac{   52.87}{ -188.64}$	&$\frac{  -16.74}{   -9.79}$	&$\frac{    5.29}{   17.08}$	&$\frac{    4.44}{   17.11}$	&$\frac{   -6.47}{   13.46}$	&$\frac{    5.43}{   15.50}$	&$\frac{   -9.91}{    3.11}$	&$\frac{    3.34}{    4.84}$	&$\frac{  -10.59}{  -10.07}$\\		
CHN	&$\frac{  195.55}{   35.76}$	&$\frac{-2324.27}{-1564.33}$	&$\frac{  106.47}{   31.54}$	&$\frac{  180.96}{   36.26}$	&$\frac{  232.75}{   40.17}$	&$\frac{  126.87}{   31.74}$	&$\frac{  -81.64}{  -33.03}$	&$\frac{ -135.91}{   -2.35}$	&$\frac{  186.27}{   38.81}$\\		
DEU	&$\frac{   38.68}{   39.59}$	&$\frac{  -10.75}{  -52.20}$	&$\frac{ -253.88}{ -139.10}$	&$\frac{  -38.50}{  -44.43}$	&$\frac{    0.50}{   -0.04}$	&$\frac{  -45.19}{  -47.50}$	&$\frac{   14.09}{    7.10}$	&$\frac{  -37.45}{ -205.30}$	&$\frac{   41.38}{   44.19}$\\		
FRA	&$\frac{   21.86}{   20.97}$	&$\frac{   -9.83}{  -22.47}$	&$\frac{  -69.45}{  -62.18}$	&$\frac{  -41.92}{  -12.66}$	&$\frac{  -17.07}{  -22.86}$	&$\frac{  -67.71}{  -56.16}$	&$\frac{    1.87}{    1.19}$	&$\frac{  -43.68}{  -97.76}$	&$\frac{   22.93}{   25.18}$\\		
GBR	&$\frac{   15.92}{   -5.81}$	&$\frac{  -12.59}{  -23.38}$	&$\frac{  -67.31}{  -54.42}$	&$\frac{  -49.89}{  -44.88}$	&$\frac{  164.52}{  145.52}$	&$\frac{  -42.47}{  -29.07}$	&$\frac{   -7.26}{  -13.58}$	&$\frac{  -33.94}{ -117.05}$	&$\frac{   17.78}{   26.54}$\\		
ITA	&$\frac{   18.26}{   19.31}$	&$\frac{    1.06}{  -15.52}$	&$\frac{  -48.34}{  -45.02}$	&$\frac{  -36.50}{  -36.91}$	&$\frac{   -1.16}{   -0.96}$	&$\frac{ -257.39}{ -190.00}$	&$\frac{    4.94}{    0.75}$	&$\frac{  -32.92}{ -135.50}$	&$\frac{   20.83}{   23.82}$\\		
JPN	&$\frac{   66.38}{  212.93}$	&$\frac{ -191.11}{ -612.51}$	&$\frac{   35.78}{  163.77}$	&$\frac{   56.09}{  212.97}$	&$\frac{   61.37}{  201.14}$	&$\frac{   46.13}{  200.28}$	&$\frac{ -726.94}{-1518.40}$	&$\frac{  -19.37}{  117.48}$	&$\frac{   69.31}{  236.68}$\\		
RUS	&$\frac{   12.12}{    7.85}$	&$\frac{  -20.29}{  -11.83}$	&$\frac{  -16.42}{    4.56}$	&$\frac{    4.27}{    7.57}$	&$\frac{    8.86}{    8.30}$	&$\frac{   -3.98}{    3.36}$	&$\frac{   -3.65}{    3.75}$	&$\frac{ -832.77}{-2139.12}$	&$\frac{   12.34}{    8.00}$\\		
USA	&$\frac{ -142.03}{ -638.28}$	&$\frac{ -567.12}{ -794.20}$	&$\frac{ -449.08}{ -239.99}$	&$\frac{ -308.90}{ -168.36}$	&$\frac{ -283.70}{ -260.03}$	&$\frac{ -279.33}{ -167.92}$	&$\frac{ -411.84}{ -742.34}$	&$\frac{ -410.17}{ -643.02}$	&$\frac{  834.90}{  640.50}$\\		
\hline \hline
\multicolumn{10}{l}{\small * The numerators are for 2017 and the denominators are for 2000.} 
\end{tabular}
}
\end{table}

Table~\ref{tbl:d_pi_Pii} presents the derivatives of $\log \pi$ with respect to $\log P_{ii}$ for 2017 and 2000, calculated using Eqs.~(\ref{eq:pi_i_2_P_ii}), (\ref{eq:pi_minus_i_2_P_ii}), and (\ref{eq:percentchange}).
The diagonal values indicates that globalization significantly benefited China, Germany, Italy, Japan, and Russia.
Canada gained in 2000 but faced negative effects in 2017.
For France, the impact was negligible due to small numerator and denominator values.
The USA and the UK appeared as anti-globalization advocates, which aligned with recent developments such as the Trump administration's withdrawal from international organizations and the UK's Brexit.
While business interest groups profited from vast international markets and low-cost raw materials,
many low-income Americans --- affected by manufacturing offshoring --- demanded cheaper goods from China.
This dynamic increased trade deficits and national debts but also delivered low inflation and affordable consumer products.
China emerged as the largest beneficiary of globalization in 2017, with a 2.324\% increase in $\pi_i$ for every 1\% decrease in $P_{ii}$. This advantage underscores the positive momentum behind its Belt and Road Initiative (BRI).
In contrast, Russia was the largest winner in 2000.

Off-diagonal values reveal spillover effects on third parties.
For example, further globalization by the USA benefited all eight other countries, suggesting that U.S. protectionist policies would likely face resistance from its partners.
China consistently gained the most from U.S. globalization in both years.
Conversely, Canada experienced minimal impact from U.S. protectionism in 2017, and its own globalization would have only marginally benefited the USA, as indicated in the first row.
The last column shows that additional globalization from seven other countries would negatively affect the USA, with Japan and China exerting the greatest adverse impact in 2000 and 2017, respectively.
Over 17 years, Japan's negative effect decreased by 70.2\%, while China's increased by 380\%.
Furthermore, the second row highlights the growing opposition of the USA and its Western allies to China's BRI-like projects, which intensified by a factor of $4.74$ between 2000 and 2017.

We assess the externalities imposed on third parties by the China-USA trade war during 2018 and 2019.
Table~\ref{tbl:side_effects} is derived from Eqs.~(\ref{eq:pi_i_2_P_ji}), (\ref{eq:pi_minus_i_2_P_ji}), and (\ref{eq:percentchange}).
When the USA reduced its imports from China by 1\% in 2018 and 2019, China's competitiveness $\pi_j$ suffered significant losses of .1058\% and .0937\%, respectively, while the USA's $\pi_i$ experienced non-significant gains of .0540\% and .0460\%, respectively.
Canada benefited as a free rider, whereas the other six countries were slightly negatively affected in both years. 
From 2018 to 2019, the negative effects were slightly mitigated for all countries.
Consequently, these countries were unlikely to take immediate action to prevent the dispute.
However, cumulative spillover effects are expected to be significant as the conflict is likely to persist for years.
Among potential allies, Canada and Russia would have been the best partners for the USA and China, respectively.
The second-best allies would have been the UK for the USA and Japan for China.

\begin{table}[ht]
\caption{$1000\times \frac{\mathrm{d} \log\pi_k}{\mathrm{d} \log P_{ji}}$ --- Side Effects of the China-USA Trade War on Third Parties$^*$} 
\label{tbl:side_effects}
\centering
\resizebox{.95\hsize}{!}{
\begin{tabular}{r||rrrrrrr|rr}\hline \hline
Year&CAN    &DEU   &FRA &GBR &ITA  &JPN   &RUS   &CHN    &USA    \\ \hline
2018&-26.114&7.103&3.013&.444&6.661&13.639&15.097&105.808&-53.960\\  
2019&-21.815&3.223&2.133&.039&5.324&12.392&14.894&93.744 &-45.954\\ \hline \hline
\multicolumn{10}{l}{\small * Country $i$ refers to the USA, $j$ to China, and $k$ to others.}   \\
\end{tabular}
}
\end{table}

Finally, the cooperative solution $\lambda_{ji}^*$ could mitigate the conflict and prevent further escalation.
At the inception of the conflict in 2018:
$$
\frac{\pi_j}{\pi_i} = \frac{.092618}{.284327} < \frac{g_j}{g_i} = \frac{17494.79}{23511.17}.
$$
For a positive $\Delta P_{ji}$, any $\lambda_{ji}$ between $\pi_j /\pi_i$ and $g_j/g_i$ would not only make $\frac{\mathrm{d}\log \pi_i}{\mathrm{d}\log P_{ji}}<0$ but also create further trade deficits for the USA (see Table~\ref{tbl:2scenarios}).
Thus, an economic conflict with a negative $\Delta P_{ji}$ was possible, and in this lose-lose scenario, the Nash bargaining solution was: 
$$
\lambda_{ji}^* = \sqrt{\frac{.092618\times 17494.79}{.284327 \times 23511.17}} = .49233.
$$

For each dollar decrease in exports from China to the USA ($\Delta P_{ji} = -\$ 1 / g_j$),
China would have adjusted its imports from the USA by: $\Delta P_{ij} = \lambda_{ji}^* \Delta P_{ji}$ of the USA's production, which is $.49233 * (-1/ g_j) * g_i = -\$.66164$.
Therefore, if this resolution had been applied, the USA would have earned a surplus of $.33836=(-.66164)-(-1)$ for each  dollar of reduced imports from China.
Additionally, by Theorem ~\ref{thm:trade_war}, the USA's $\pi_i$ would have increased since $\lambda^*_{ji}> \pi_j/\pi_i$ and $\Delta P_{ji}<0$.

To maintain competitive advantage and reduce national debt by earning trade surpluses, the USA, by the impossibility trilemma, had to reduce its imports and exports --- i.e., $\Delta P_{ji}<0$ --- through policies such as tariffs and manufacturing reshoring.

\section{Policy Implications and Discussions}\label{sect:discussions}

Economic globalization and trade wars have become pressing public concerns, driven by intensifying global competition, large-scale labor market restructuring, and the international distribution of added value.
This study argues that import and export data already capture these dynamics and that competitive advantage --- rather than comparative advantage --- is the primary force behind trade wars and the trend toward deglobalization.

Our objective is to extract insights into a nation's competitive strength and bargaining power from trade data, identify optimal strategies for managing trade conflicts and fostering cooperation, and assess the appropriate level of engagement in globalization. 
When applying this policy-oriented framework in real-world trade scenarios, it is essential to explore alternative approaches that adapt and extend the framework to specific national and economic contexts.

\subsection{Practical and Policy Implications}
This paper conceptualizes national competitiveness as a systemic property, rather than a simple reflection of unilateral production metrics or bilateral trade balances.
Specifically, the metric $\pi$ represents stationary standings among nations, derived from their economic interactions --- namely, bilateral trade flows. 
This formulation positions $\pi$ as a network-wide measure of economic strength, accounting for both direct and indirect effects across global value chains. 
Policymakers should monitor $\pi$ and its fluctuations alongside trade balances and production metrics to gain a more comprehensive understanding of national competitiveness.

\subsubsection{Prioritizing Trade Objectives}
When navigating the trade-off between competitiveness and trade balance, two actionable ratios can guide bilateral trade decisions:
long-run competitiveness-neutral point $\lambda_{ji} = \dfrac{\pi_j}{\pi_i}$,  useful for managing strategic rivalries without escalation;
short-run trade-balance-neutral point $\lambda_{ji} = \dfrac{g_j}{g_i}$, suitable for addressing immediate concerns such as employment, debt sustainability, or foreign exchange reserves.
These benchmarks help trade negotiators align policy actions with either long-term strategic goals or short-term economic stability. 

The choice between strategies based on competitive versus comparative advantage should reflect national characteristics.
Small economies are advised to adopt comparative advantage strategies to promote inclusive growth and mutual benefit, while superpowers may pursue competitive advantage strategies that favor dominant players.
Most countries will adopt mixed, country-specific objectives. For example, in 2020, the USA faced strategic challenges from China alongside massive national debt.

Yet, competitiveness is often overlooked by incumbent administrations due to its intangible nature and long-term impact.
Few quantitative metrics like $\pi$ have been used to measure competitiveness growth.

\subsubsection{Recognizing Asymmetries Early}
The convex nature of competitiveness creates reinforcing dynamics --- often referred to as the Matthew effect --- which makes it difficult for latecomers to catch up unless they grow substantially faster. 
This has critical implications for strategic sequencing, such as determining when to confront or collaborate with emerging rivals.
Convexity also increases the risk of a Thucydides trap, where instability arises from competition between the two leading superpowers. Stability is more likely when a single superpower coexists with several great powers.
Our empirical studies suggest that such instability may persist due to ongoing competition between China and the USA.

Consequently, incumbent superpowers should prioritize maintaining their leadership position.
Meanwhile, small, open economies can leverage agglomeration effects by aligning with high-$\pi_i$ hubs and using pricing strategies to offset biases toward large markets.
Fast-growing challengers aiming to close the competitiveness gap should avoid systematic confrontations with significantly larger economies.

Another critical asymmetry is the impossibility trilemma, which clarifies trade-offs often obscured in policy debates.
Large economies should use it to set realistic bilateral goals, recognizing that achieving both a surplus and higher competitiveness may require reducing exports to challengers or broadly curbing imports.

\subsubsection{Competition and Cooperation Corridors} 
The decision to compete or collaborate with a specific country depends on the marginal impact of imports and exports on $\pi_i$.
Given limited resources, country $i$ may prioritize its top competitors and collaborators while ignoring others.

The counterbalance equilibrium provides a formal rationale for mixed trade policies that combine competitive pressure with cooperative incentives. This unity of opposites suggests avoiding extreme actions in both retaliations and cooperation.
For example, controlling rare minerals and advanced technologies could escalate issues to national security levels.

Trade relationships evolve over time, making timing critical. Countries should develop an early-warning dashboard that
annually tracks changes in $\pi$, $\beta$, and derivatives in Theorems~\ref{thm:trade_war} and \ref{thm:globalization}. 
Thresholds --- such as elasticities exceeding $\pm .1\%$ --- should trigger interagency reviews if signs flip or magnitudes shift significantly.

\subsubsection{Negotiation leverage and institutional design} 
The square-root formulation in $\lambda_{ji}^*$ symmetrically balances competitiveness and trade balance.
It is scale-free and can serve as a default ``fair retaliation'' coefficient in bilateral trade protocols.
However, this solution can be strategically manipulated (Corollary~\ref{cr:homogeneous_preference}), particularly if a weaker nation appears to avoid competition with a dominant one, which may overlook this trap. The asymmetric Nash solution in Eq.~(\ref{eq:aNash}) also benefits smaller countries more than the non-cooperative solutions in Corollary~\ref{cr:Nash_confrontation}.

Moreover, when analyzing the relationship between a weaker country $j$ and a dominant country $i$, the Matthew effect indicates that the Nash bargaining solution yields $\lambda_{j i}^*>\pi_j / \pi_i$. Consequently, the United States may identify only a limited number of countries as suitable partners for deeper collaboration in its pursuit of enhanced national competitiveness (see Section~\ref{sect:identify_competitor_collaborator}). Nevertheless, all other economies can benefit from cooperating with the sole superpower, as outlined in Corollary~\ref{cr:agglomeration} and supported by the empirical study.

Both the bilateral solution $\lambda_{ji}^*$ and the multilateral solution $\beta_i$ faces drawbacks due to simplifying assumptions that often fail to capture the complexity of international relations.
Incorporating unequal bargaining power and third-party influence can improve these solutions.
For instance, the USA may impose stricter retaliation measures (e.g., tariffs and quotas) 
on nations running trade surpluses with it. This may exclude China, given their comparable production sizes, with or without purchasing power parity.

\subsubsection{Globalization and Supply-Chain strategy} 
Before adjusting border measures, country $i$ should assess its sensitivity to changes in its domestic share $P_{ii}$,
i.e., $\dfrac{\mathrm{d} \pi_i}{\mathrm{d} P_{ii}}$. This indicates whether further globalization or protectionism enhances competitiveness.
This empirical approach offers a pragmatic alternative to ideological debates over liberation versus reshoring.

A mixed strategy appears optimal: deepen ties with partners in $\mathcal{N}_i^+$ and selectively ring-fence against those in $\mathcal{N}_i^-$. In practice, however, both sets are too large.
Accordingly, country $i$ should consider joining or upgrading PTAs/RTAs rich in $\mathcal{N}_i^+$, while negotiating carve-outs for areas with high $\mathcal{N}_i^-$ exposure --- such as rules-of-origin thresholds or snap-back clauses. With these agreements, the spillover effects in Eqs.~(\ref{eq:pi_minus_i_2_P_ji}) and (\ref{eq:pi_minus_i_2_P_ii}) require adjustments.

Due to the 2018 China-USA trade war, fragmentation between two blocs of nations has increased tension in the international trade system. The first bloc includes the over-globalized USA and its traditional allies, while the second comprises nations heavily reliant on China's infrastructure-building capacity and affordable goods.
Trade between blocs may decrease, while intra-bloc trade may increase.
Countries outside these blocs may benefit.

\subsubsection{Third-party spillovers and Trade Alliances}
Partners play a pivotal role in the trade network. While uniform trade policies rarely benefit all partners, heterogeneous policies can generate gains across countries.
Several methods --- such as Eqs.~(\ref{eq:pi_i_pi_j_breakeven}) and (\ref{eq:percentchange}), or the sets $\mathcal{N}^+_i$ and $\mathcal{N}^-_i$ --- help identify competitors and collaborators.

Given any partner, country $i$ can enhance its competitiveness by adopting targeted strategies: weakening ties with competitors or strengthening ties with collaborators. 
These strategies adjust existing trade relationships, and
before implementing them, a country should anticipate who stands to gain and who might resist --- critical for coalition-building and mitigating opposition.
 
Gains and losses extend beyond principal actors to third-party countries. 
To manage these dynamics, countries should use
Eqs.~(\ref{eq:pi_minus_i_2_P_ji}) and (\ref{eq:pi_minus_i_2_P_ii}) to measure spillover effects, forecast ally reactions, and design compensatory mechanisms --- thereby reducing coalition leakage during disputes.
Ignoring such side effects can turn collaborators into competitors.
Partner-specific collaboration can yield substantial returns, but these benefits often diminish over time, underscoring the need to regularly update partner rankings and reassess traditional alliances using current data.

\subsection{Discussions and Extensions}
The methodologies in this study --- rooted in network and game theory --- emphasize real-time interactions among economies without relying on fixed parameters or predefined utility functions.
These network game approaches can also be applied to other bilateral or dyadic systems, such as traffic networks, social platforms, and global payment infrastructures, as well as analogous conflicts, including currency and technology wars.

Several critical issues warrant further exploration.
First, significant trade imbalances can create counterbalance disequilibrium.
For example, when country $i$ uses borrowed funds to purchase exports from country $j$, $i$'s consumption still  influence $j$'s production.
However, this influence originates not only from $i$ but also from the lender, who exerts some control over $i$'s consumption.
If the lender is country $j$ itself, then $\pi_i$ may be overestimated and $\pi_j$ underestimated, potentially leading to a debt trap for $i$.

Second, unlike economic superpowers, emerging economies may prioritize export growth over competitive advantages.
To close the competitiveness gap, their economic growth must significantly outpace that of advanced economies, consistent with the Matthew effect in $\pi$.
Consequently, maximizing competitive advantages may be a low priority.
For instance, during China's early capital accumulation in the 1980s and 1990s, earning additional U.S. dollars was crucial to finance industrial modernization, such as high-speed rail technology transfer.
Incorporating the time dimension --- currently absent from this research --- is essential for developing effective growth strategies.

Thirdly, key dimensions of economic globalization --- namely capital, information, and technology --- remain insufficiently integrated into the analysis. 
Although these elements interact with goods and services, their associated data are comparatively less accessible.
Additionally, several data limitations may hinder empirical assessments, including missing values in service sectors and frequent discrepancies between country $i$'s reported imports from country $j$ and $j$'s reported exports to $i$.

The integration of regional economies also warrants deeper investigation.
Theorem~\ref{thm:trade_war} and $\mathcal{N}_i^+$ could be employed to identify suitable candidates for integration;
however, the underlying assumptions may require revision.
For example, if country $i$ participates in multiple RTAs, then $\Delta P_{ii}$ may not be proportionally offset by other countries, as assumed in Theorem~\ref{thm:globalization}.

Moreover, additional variables could be incorporated into the demand-side counterbalance, particularly from the supply side --- such as labor mitigation --- which could refine the fractions in $P$. 
For instance, accounting for labor loss in $g_i$ would capture factors directly influencing country $i$'s production.
These factors could be reflected in the $i$th row of $P$.
While consumption may still predominantly drive these fractions, not all production is consumption-driven.

The framework offers several avenues for extension.
First, in the context of a bilateral trade war and rising anti-globalization sentiments, a country may target specific groups of partners without directly affecting others.
For example, as shown in Table~\ref{tbl:d_pi_pji}, the USA could simultaneously raise tariffs on imports from China, Germany, and Japan.

Second, while the dynamical systems described by Eqs.~(\ref{eq:pi_i_2_P_ji})--(\ref{eq:pi_minus_i_2_P_ji}) or (\ref{eq:pi_i_2_P_ii})--(\ref{eq:pi_minus_i_2_P_ii}) assume a fixed $P$, introducing a temporal dimension could yield valuable insights.
Incorporating a time horizon would enable analysis of a trade war's effects on inflation, $\pi$'s dependence on historical values, and improvements in trade deficits.
This temporal extension could also enhance the model's forecasting capabilities.
Intertemporal delays often arise when production and consumption are distributed across multiple years.
For instance, if production spans two years, imports of intermediate goods from the previous year could be included in the current year's completed production.

Next, we could allow $\lambda_{ji}$ to depend on additional determinants beyond competitiveness and production ratios.
Industrial or geopolitical analyses may guide this choice.
For instance, in 2000, Japan and South Korea were competitors in automobile and electronic exports, making $\pi_j/\pi_i$ a suitable basis for $\lambda_{ji}$.
In contrast, no major industry served as a battleground between China and the USA that year, suggesting that $g_j/g_i$ or $\lambda_{ji}^*$ might be more appropriate.

The literature (e.g., Allen, Arkolakis, and Takahashi, 2020; Isard, 1954; Yilmazkuday, 2021) highlights the significance of distance between trade partners and the type of merchandise as key determinants of trade flows.
Accordingly, one might discount the $i$th row of $P$ by the distances between country $i$ and its trade partners, thereby partially removing locational advantages from $\pi$.

Econometric analyses could also offer further insights. 
For example, in the linear regression of Eq.~(\ref{eq:pi_g_gsq}), a positive or negative estimated residual $\hat \epsilon_i$ may indicate that $\pi_i$ is overvalued or undervalued, respectively.
Including additional regressors, such as the debt-to-GDP ratio, could enhance the explanatory power of $\pi$.

Finally, in anticipation of the side effects of a trade war between country $i$ and country $j$, the latter might seek cooperation with third parties that would also be adversely affected.
Country $i$ could also form an alliance with those who would benefit from the conflict.
However, expanding the number of countries involved on both sides could escalate the situation into a global trade war, reminiscent of the Cold War era (1947--1991).





\section*{Appendix}

\subsection*{A1. Proof of Theorem~\ref{thm:uniqueness_pi}}

\noindent 
We multiply $\vec{1}_n$ from the right on both sides of the equation $\rho P = c \rho$ to get 
$\rho P \vec{1}_n = c \rho \vec{1}_n.$
Using $P \vec{1}_n=\vec{1}_n$ and $\rho \vec{1}_n=1$, we obtain
$\rho \vec{1}_n = c.$
Thus, $c=1$.
Also, by $\rho P = c \rho = \rho$ and the uniqueness of $\pi$, we have $\rho=\pi$.

\subsection*{A2. Proof of Corollary~\ref{cr:agglomeration}}
Assume that $g_j>g_k$.
For example, country $i$ considers certain imports from either country $j$ or $k$, all else being equal. 
If $i$ selects $j$, by Eq.~(\ref{eq:power_in}), then the competitiveness country $i$ derives from $j$ in the imports is approximately
$$
\pi_j \Delta P_{ji} = \pi_j \times \frac{\mathrm{the}\ \mathrm{imports}\ \mathrm{from}\ j}{g_j} = \frac{\pi_j}{g_j} \times (\mathrm{the}\ \mathrm{imports}\ \mathrm{from}\ j)
$$
where $\Delta P_{ji}$ is the change on $P_{ji}$ due to the imports.
The approximation ignores the ripple effect from the shock.
If $i$ selects $k$, then the competitiveness country $i$ derives from $k$ in the imports is approximately
$$
\pi_k \Delta P_{ki} = \pi_k \times \frac{\mathrm{the}\ \mathrm{imports}\ \mathrm{from}\ k}{g_k} = \frac{\pi_k}{g_k} \times (\mathrm{the}\ \mathrm{imports}\ \mathrm{from}\ k).
$$
Because of the Matthew effect, ${\pi_j}/{g_j}>{\pi_k}/{g_k}$, and so, country $i$ would likely choose $j$, all else being equal.

\subsection*{A3. Proof of Theorem~\ref{thm:trade_war}}
We apply matrix calculus with the restrictions of Eq.~(\ref{eq:equilibrium}) and $\pi \vec{1}_n=1$ to prove the theorem.
When making a small shock or perturbation $\Delta P$ to $P$ in Eq.~(\ref{eq:equilibrium}), the new authority distribution $\pi + \Delta \pi$ satisfies the counterbalance equation:
\begin{equation}\label{eq:delta_authority_distribution}\tag{A.1}
\pi + \Delta \pi = (\pi + \Delta \pi) [P + \Delta P]
\end{equation}
subject to $\Delta P \vec{1}_n = \vec{0}_n$ and $\Delta \pi \vec{1}_n = 0$.
After subtracting $\pi = \pi P$ from Eq.~(\ref{eq:delta_authority_distribution}), we get
$
\Delta \pi [I_n - P] = \pi \Delta P + \Delta \pi \Delta P
$
and its first-order approximation is:
\begin{equation}\label{eq:linear_approximation}\tag{A.2}
\Delta \pi [I_n - P] \approx \pi \Delta P.
\end{equation}
 
We let $P_{ji}$ have a small change $\Delta P_{ji}$ and attempt to calculate its effect on $\pi$, including $\pi_i$, $\pi_{-i}$, and $\pi_j$. 
Accordingly, $P$ has three other simultaneous changes: $-\Delta P_{ji}$ on $P_{jj}$ to offset the change on $P_{ji}$ in the $j$th row;
$\lambda_{ji} \Delta P_{ji}$ on $P_{ij}$ for country $j$'s retaliation upon $i$'s change at $P_{ji}$; 
and $-\lambda_{ji} \Delta P_{ji}$ on $P_{ii}$ to maintain the unit sum of the $i$th row. 
Therefore, $\frac{\mathrm{d} P}{\mathrm{d} P_{ji}} = \lim\limits_{\Delta P_{ji}\to 0} \frac{\Delta P}{\Delta P_{ji}}$ is a zero $n\times n$ matrix except for:
$1$ at $(j,i)$; $-1$ at $(j,j)$; $\lambda_{ji}$ at $(i,j)$; and $-\lambda_{ji}$ at $(i,i)$.
Without loss of generality, we assume $1\le i < j\le n$. 
Dividing Eq.~(\ref{eq:linear_approximation}) by $\Delta P_{ji}$ and letting $\Delta P_{ji} \to 0$, we get the following equation for the derivative of $\pi$ with respect to $P_{ji}$:
\begin{equation}\label{eq:partial_pi}\tag{A.3}
\frac{\mathrm{d} \pi}{\mathrm{d} P_{ji}} \left [ I_n -P \right ] 
= 
\pi \frac{\mathrm{d} P}{\mathrm{d} P_{ji}} 
= 
\left (\vec{0}^\top_{i-1}, -\lambda_{ji}\pi_i+\pi_j, \vec{0}^\top_{j-i-1}, \lambda_{ji}\pi_i-\pi_j, \vec{0}^\top_{n-j} \right )
\end{equation}
where the right-hand vector has all zeros except for the $i$th and the $j$th elements.

Also, we partition the transpose of $P$ as 
\begin{equation}\label{eq:partition_of_P}\tag{A.4}
P^\top = \left [
\begin{array}{ccccc}
H_1   &\eta_1        &H_2  &\eta_2        &H_3   \\
\mu_1 &\boxed{P_{ii}}&\mu_2&\boxed{P_{ji}}&\mu_3 \\
H_4   &\eta_3        &H_5  &\eta_4        &H_6   \\
\mu_4 &\boxed{P_{ij}}&\mu_5&\boxed{P_{jj}}&\mu_6 \\
H_7   &\eta_5        &H_8  &\eta_6        &H_9   
\end{array}
\right ]
\end{equation}
where $\eta_1,\cdots,\eta_6$ are column vectors, $\mu_1,\cdots,\mu_6$ are row vectors, and $H_1,\cdots, H_9$ are sub-matrices of $P^\top$.
We write the augmented matrix for the identity $\frac{\mathrm{d} \pi}{\mathrm{d} P_{ji}} \vec{1}_n = 0$ and the transpose of Eq.~(\ref{eq:partial_pi}) as
\begin{equation}\label{eq:augment_ij} \tag{A.5}
\resizebox{.9\hsize}{!}{$
\left [
\begin{array}{ccccc|c}
\vec{1}_{i-1}^\top&1               &\vec{1}_{j-i-1}^\top&1               &\vec{1}_{n-j}^\top&0\\
I_{i-1}-H_1       &-\eta_1         &-H_2                &-\eta_2         &-H_3              &\vec{0}_{i-1}  \\
-\mu_1            &\boxed{1-P_{ii}}&-\mu_2              &\boxed{-P_{ji}} &-\mu_3            &-\lambda_{ji}\pi_i+\pi_j\\
-H_4              &-\eta_3         &I_{j-i-1}-H_5       &-\eta_4         &-H_6              &\vec{0}_{j-i-1}\\
-\mu_4            &\boxed{-P_{ij}} &-\mu_5              &\boxed{1-P_{jj}}&-\mu_6            &\lambda_{ji}\pi_i-\pi_j\\
-H_7              &-\eta_5         &-H_8                &-\eta_6         &I_{n-j}-H_9       &\vec{0}_{n-j}
\end{array}
\right ].
$}
\end{equation}

Since all rows of Eq~(\ref{eq:augment_ij}), except for the first one, sum up to a zero vector, we add them to the $(i+1)$st row, making
the $(i+1)$st row of Eq.~(\ref{eq:augment_ij}) a zero vector. 
After dropping the $(i+1)$st row in Eq.~(\ref{eq:augment_ij}) and moving the $i$th column to the first column without changing the order of other columns, we obtain the augmented matrix for $\left (\frac{\mathrm{d} \pi_i}{\mathrm{d} P_{ji}}, \frac{\mathrm{d} \pi_{-i}^\top}{\mathrm{d} P_{ji}} \right )^\top$:
\begin{equation}\label{eq:simplied_augmented}\tag{A.6}
\left[ 
\begin{array}{cc | c}
1        & \vec{1}_{n-1}^\top& 0 \\
-\alpha_i& I_{n-1}-Z_i       & (\lambda_{ji}\pi_i-\pi_j)\gamma_{ji} \\ 
\end{array}
\right].
\end{equation}

To make the matrix in Eq.~(\ref{eq:simplied_augmented}) lower-triangular, we multiply the following non-singular matrix
\begin{equation}\label{eq:left_multiplication}\tag{A.7}
\left [
\begin{array}{cc}
1            &-\vec{1}_{n-1}^\top (I_{n-1}-Z_i)^{-1} \\
\vec{0}_{n-1}&I_{n-1}
\end{array}
\right ]
\end{equation}
to the left side of Eq.~(\ref{eq:simplied_augmented}) to get:
\begin{equation}\label{eq:trans_augmented}\tag{A.8}
\resizebox{.85\hsize}{!}{$
\left[ 
\begin{array}{cc | c}
1+\vec{1}_{n-1}^\top (I_{n-1}-Z_i)^{-1} \alpha_i&\vec{0}_{n-1}^\top&-(\lambda_{ji}\pi_i-\pi_j)\vec{1}^\top_{n-1} (I_{n-1} -Z_i)^{-1} \gamma_{ji} \\
-\alpha_i                                       &I_{n-1} -Z_i      &(\lambda_{ji}\pi_i-\pi_j)\gamma_{ji} \\ 
\end{array}
\right].
$}
\end{equation}
Therefore, by the first row of Eq.~(\ref{eq:trans_augmented}):
$$
\frac{\mathrm{d} \pi_i}{\mathrm{d} P_{ji}} 
= 
-\frac{(\lambda_{ji}\pi_i-\pi_j)\vec{1}^\top_{n-1} (I_{n-1} -Z_i)^{-1} \gamma_{ji}}{1+\vec{1}_{n-1}^\top (I_{n-1}-Z_i)^{-1} \alpha_i}.
$$
Also, by the second row of Eq.~(\ref{eq:trans_augmented}):
$$
-\frac{\mathrm{d} \pi_i}{\mathrm{d} P_{ji}} \alpha_i + \left (I_{n-1}-Z_i \right ) \frac{\mathrm{d} \pi_{_{-i}}}{\mathrm{d} P_{ji}} 
= 
(\lambda_{ji} \pi_i - \pi_j) \gamma_{ji}
$$
and thus:
$$
\begin{array}{rcl}
\frac{\mathrm{d} \pi_{_{-i}}}{\mathrm{d} P_{ji}}
&=& 
\left (I_{n-1}-Z_i \right )^{-1} \left [
(\lambda_{ji} \pi_i - \pi_j) \gamma_{ji}
+ 
\frac{\mathrm{d} \pi_i}{\mathrm{d} P_{ji}} \alpha_i  
\right ] \\
&=&
(\lambda_{ji}\pi_i-\pi_j)
\left (I_{n-1}-Z_i \right )^{-1} \left [ 
\gamma_{ji}
-
\frac{\vec{1}^\top_{n-1} (I_{n-1} -Z_i)^{-1} \gamma_{ji}}{1+\vec{1}_{n-1}^\top (I_{n-1}-Z_i)^{-1} \alpha_i}
\alpha_i
\right ].
\end{array}
$$

Similarly, after we add all rows (except for the first one) in Eq.~(\ref{eq:augment_ij}) to the $(j+1)$st row, the $(j+1)$st row becomes a zero vector. 
After dropping the $(j+1)$st row in Eq.~(\ref{eq:augment_ij}) and moving the $j$th column to the first column without changing the order of other columns, we obtain the augmented matrix for $\left (\frac{\mathrm{d} \pi_j}{\mathrm{d} P_{ji}}, \frac{\mathrm{d} \pi_{-j}^\top}{\mathrm{d} P_{ji}} \right )^\top$:
\begin{equation}\label{eq:simplied_augmented_j}\tag{A.9}
\left[ 
\begin{array}{cc | c}
1        &\vec{1}_{n-1}^\top&0 \\
-\alpha_j&I_{n-1}-Z_j       &(-\lambda_{ji}\pi_i+\pi_j)\gamma_{ij} \\ 
\end{array}
\right].
\end{equation}
We multiply the following non-singular matrix:
$$
\left [
\begin{array}{cc}
1            &-\vec{1}_{n-1}^\top (I_{n-1}-Z_j)^{-1} \\
\vec{0}_{n-1}&I_{n-1}
\end{array}
\right ]
$$
to the left side of Eq.~(\ref{eq:simplied_augmented_j}) to get:
\begin{equation}\label{eq:trans_augmented_j}\tag{A.10}
\resizebox{.85\hsize}{!}{$
\left[ 
\begin{array}{cc | c}
1+\vec{1}_{n-1}^\top (I_{n-1}-Z_j)^{-1} \alpha_j&\vec{0}_{n-1}^\top&(\lambda_{ji}\pi_i-\pi_j)\vec{1}^\top_{n-1} (I_{n-1} -Z_j)^{-1} \gamma_{ij} \\
-\alpha_j                                       &I_{n-1}-Z_j       &(-\lambda_{ji}\pi_i+\pi_j)\gamma_{ij} \\ 
\end{array}
\right].
$}
\end{equation}
Therefore, by the first row of Eq.~(\ref{eq:trans_augmented_j}), 
$$
\frac{\mathrm{d} \pi_j}{\mathrm{d} P_{ji}}
= 
\frac{(\lambda_{ji}\pi_i-\pi_j)\vec{1}^\top_{n-1} (I_{n-1} -Z_j)^{-1} \gamma_{ij}}{1+\vec{1}_{n-1}^\top (I_{n-1}-Z_j)^{-1} \alpha_j}.
$$

\subsection*{A4. Proof of Corollary~\ref{cor:opposite_sign}}

First, both
$
(I_{n-1}-Z_i)^{-1}=I_{n-1}+Z_i+Z_i^2+Z_i^3+\cdots
$
and
$
(I_{n-1}-Z_j)^{-1}=I_{n-1}+Z_j+Z_j^2+Z_j^3+\cdots
$
are non-negative matrices.
Secondly, $\vec{1}_{n-1}$, $\gamma_{ji}$, $\gamma_{ij}$, $\alpha_i$, and $\alpha_j$ are all non-negative.
Therefore, by Eqs.~(\ref{eq:pi_i_2_P_ji}) and (\ref{eq:pi_j_2_P_ji}) in Theorem~\ref{thm:trade_war}, $\frac{\mathrm{d}\pi_i}{\mathrm{d} P_{ji}} \frac{\mathrm{d}\pi_j}{\mathrm{d} P_{ji}} \le 0$.
Also, 
$$
\frac{\mathrm{d}\pi_i}{\mathrm{d} P_{ij}} \frac{\mathrm{d}\pi_j}{\mathrm{d} P_{ij}} 
=
\frac{\mathrm{d}\pi_i}{\lambda_{ji} \mathrm{d} P_{ji}} \frac{\mathrm{d}\pi_j}{\lambda_{ji} \mathrm{d} P_{ji}} 
=
\frac{1}{\lambda_{ji}^2} \frac{\mathrm{d}\pi_i}{\mathrm{d} P_{ji}} \frac{\mathrm{d}\pi_j}{\mathrm{d} P_{ji}} 
\le
0.
$$

\subsection*{A5. Proof of Theorem~\ref{thm:zero_trade_deficit}}

In added value, the changed exports from country $j$ to $i$ are $g_j\Delta P_{ji}$, while those from country $i$ to $j$ are $g_i\Delta P_{ij} = g_i\lambda_{ji} \Delta P_{ji}$.
If $\lambda_{ji}=g_j/g_i$, then $g_j\Delta P_{ji} = g_i\Delta P_{ij}$, resulting in zero net trade surplus and deficit between these countries.
This condition is for instant balance change.

Over a specific time period, there is no trade surplus or deficit if and only if $g_j/g_i= P_{ij}/P_{ji}$, because
$g_i P_{ij}$ represents exports from $i$ to $j$ and $g_j P_{ji}$ represents exports from $j$ to $i$.

\subsection*{A6. Proof of Theorem~\ref{thm:impossibility_trilemma}}

When $\pi_j/\pi_i < g_j/g_i$, there are three possible intervals and two values for $\lambda_{ji}$: $0<\lambda_{ji}<\pi_j/\pi_i$, $\lambda_{ji}=\pi_j/\pi_i$,
$\pi_j/\pi_i <\lambda_{ji}< g_j/g_i$, $\lambda_{ji}=g_j/g_i$, or $g_j/g_i<\lambda_{ji}<\infty$.
The small change $\Delta P_{ji}$ could be positive or negative, and $\Delta P_{ij}=\lambda_{ji}\Delta P_{ji}$ has the same sign.
Using Theorems~\ref{thm:trade_war} and \ref{thm:zero_trade_deficit}, Table~\ref{tbl:2scenarios} lists all possible outcomes for these ten scenarios of $(\lambda_{ji}, \Delta P_{ji})$.

\begin{table}[ht]
\centering
\caption{Scenarios of $(\lambda_{ji}, \Delta P_{ji})$ and their impacts on country $i$'s $\pi_i$ and net trade with $j$$^*$}
\label{tbl:2scenarios}
\resizebox{.9\hsize}{!}{
\begin{tabular}{ r || @{\hspace{.1cm}}c@{\hspace{.1cm}} | @{\hspace{.5cm}}c@{\hspace{.5cm}} | @{\hspace{.1cm}} c @{\hspace{.1cm}} | @{\hspace{.5cm}}c@{\hspace{.5cm}} | @{\hspace{.1cm}}c@{\hspace{.1cm}} } \hline
	&$0<\lambda_{ji}<\frac{\pi_j}{\pi_i}$&$\lambda_{ji}=\frac{\pi_j}{\pi_i}$ &$\frac{\pi_j}{\pi_i} <\lambda_{ji}< \frac{g_j}{g_i}$&$\lambda_{ji}=\frac{g_j}{g_i}$&$\frac{g_j}{g_i}<\lambda_{ji}<\infty$ \\ \hline \hline
\multirow{2}{*}{$\Delta P_{ji}>0$}&$\pi_i\uparrow$ &$\Delta \pi_i=0$ &$\pi_i\downarrow$&$\pi_i\downarrow\downarrow$&$\pi_i \downarrow\downarrow\downarrow$\\ 
	&$---$ &$--$&$-$          &zero net&$+$ \\ \hline
\multirow{2}{*}{$\Delta P_{ji}<0$}&$\pi_i\downarrow$ &$\Delta \pi_i=0$ &$\pi_i\uparrow$ &$\pi_i\uparrow\uparrow$ &$\pi_i \uparrow\uparrow\uparrow$  \\ 
	&$+++$\hspace{.8cm}&$++$&$+$   &zero net&$-$       \\ \hline
\multicolumn{6}{l}{$^*${\small $+$ and $-$ for net trade surplus and deficit, respectively.}} \\
\multicolumn{6}{l}{$^*${\small Magnitudes in the same row are indicated by the numbers of $\downarrow$, $\uparrow$, $+$, or $-$.}} \\
\multicolumn{6}{l}{$^*${\small $\Delta \pi_j$ and $\Delta \pi_i$ have opposite signs and country $i$ and $j$'s net trade balances have opposite signs.}} \\
\end{tabular}
}
\end{table}

From the first row, country $i$ could not increase $\pi_i$ and achieve a trade surplus with country $j$ at the same time.
When $\pi_j / \pi_i <\lambda_{ji}< g_j/g_i$ in the same row, however, country $j$ has a trade surplus, an increasing $\pi_j$, and $\Delta P_{ji}>0$.
In these ten scenarios, none strictly dominates the others, regarding competitiveness, net trade balance, comparative advantages, and their magnitudes of changes.
On the other hand, countries $i$ and $j$ may not completely contradict in all these interests.

\subsection*{A7. Proof of Theorem~\ref{thm:Nash_bargaining}}
In the first scenario, country $j$ prefers $q$ to $p$, whereas country $i$ prefers $1/p$ to $1/q$. 
Therefore, $u_j(x)=x$ is a utility function for $j$ because $u_j(q) > u_j(p)$.
Similarly, $u_i(x)=x$ is a utility function for $i$.
These are von Neumann-Morgenstern utilities, uniquely determined up to a positive affine transformation.
Assuming that all countries have equal prior bargaining power, the Nash bargaining solution solves the following maximization problem:
\begin{equation}\label{eq:Nash}\tag{A.11}
\max\limits_{\lambda_{ji}\lambda_{ij}=1; \ \lambda_{ji}>0} \left (\lambda_{ji}-p \right ) \left (\lambda_{ij} - \frac{1}{q} \right ).
\end{equation}
In Eq.~(\ref{eq:Nash}), $\lambda_{ji}-p$ and $\lambda_{ij} - \frac{1}{q}$ are the excessive payoffs or welfare for countries $j$ and $i$ in terms of their respective utility functions.
The bargaining solution maximizes the product of the excessive utilities.

In the second scenario of opposite preferences, country $j$ prefers $p$ to $q$ while country $i$ prefers $1/q$ to $1/p$. 
Their utility functions could be $u_j(x) = u_i(x) = -x$, and their status quo points are $u_j(q)=-q$ and $u_i(1/p)=-1/p$, respectively.
The maximization problem becomes:
\begin{equation}\label{eq:Nash_threat}\tag{A.12}
\max\limits_{\lambda_{ji}\lambda_{ij}=1; \ \lambda_{ji}>0} \left (-\lambda_{ji}+q \right ) \left (-\lambda_{ij}+\frac{1}{p} \right ).
\end{equation}
In this utility function, $\lambda_{ji}$ seems to have no meaning of reprisal.
However, we can rephrase Eq.~(\ref{eq:Nash_threat}) by:
\begin{equation}\label{eq:max_loss}\tag{A.13}
\max\limits_{\lambda_{ji}\lambda_{ij}=1; \ \lambda_{ji}>0} \left(\lambda_{ji} - q \right) \left(\lambda_{ij} - \frac{1}{p}\right)
\end{equation}
in which $\lambda_{ji} - q$ and $\lambda_{ij} - 1/p$ are the loss functions for $j$ and $i$, respectively.
To coerce an agreement formation, each country plays a threat strategy to maximize its counterpart's loss.
The final outcome is Eq.~(\ref{eq:max_loss}).
Moving from $p$ to $q$, a larger $\lambda_{ji}$ means less loss to $j$; thus, $\lambda_{ji}$ still preserves the meaning of retaliation for $j$ in Eq.~(\ref{eq:max_loss}) and, so, in Eq.~(\ref{eq:Nash_threat}).

Let $x=\lambda_{ji}$. 
Then $\lambda_{ij} = 1/x$. 
By Eq.~(\ref{eq:Nash}), the Nash bargaining solution is:
$$
\argmax_{x>0} \left (x- p \right ) \left (\frac{1}{x}-\frac{1}{q} \right )
=
\argmin_{x>0} \left (\frac{x}{q} + \frac{p}{x} \right ) 
=
\sqrt{pq}
=
\sqrt{\frac{\pi_j g_j}{\pi_i g_i}}.
$$
Also, by Eq.~(\ref{eq:Nash_threat}) or (\ref{eq:max_loss}):
$$
\argmax_{x>0} \left (x- q \right ) \left (\frac{1}{x}-\frac{1}{p} \right )
=
\argmin_{x>0} \left (\frac{x}{p} + \frac{q}{x} \right ) 
=
\sqrt{pq}
=
\sqrt{\frac{\pi_j g_j}{\pi_i g_i}}.
$$

\subsection*{A8. Proof of Corollary~\ref{cr:Nash_bargaining}}

When we extend the effective bargaining set of $\lambda_{ji}$ from $[p,q]$ to $[cp, q/c]$,
the corresponding range for $\lambda_{ij}$ then changes from $[1/q,1/p]$ to $[c/q, 1/(cp)]$.
Letting $x=\lambda_{ji}$, the bargaining solution to Eq.~(\ref{eq:Nash}) becomes:
$$
\argmax_{x > 0} \left (x- cp \right ) \left (\frac{1}{x}-\frac{c}{q} \right )
=
\argmin_{x > 0} \left (\frac{cx}{q} + \frac{cp}{x} \right ) 
=
\sqrt{pq}
=
\sqrt{\frac{\pi_j g_j}{\pi_i g_i}}.
$$
In the above, we need $c>0$ and $cp\le q/c$, i.e., $0< c \le \sqrt{q/p}$.

After we replace $p$ and $q$ with $cp$ and $q/c$, respectively, Eq.~(\ref{eq:Nash_threat}) or (\ref{eq:max_loss}) becomes:
$$
\argmax_{x >0} \left (\frac{q}{c}-x \right ) \left (\frac{1}{cp} - \frac{1}{x} \right )
=
\argmin_{x > 0} \left (\frac{q}{cx} + \frac{x}{cp} \right ) 
=
\sqrt{pq}
=
\sqrt{\frac{\pi_j g_j}{\pi_i g_i}}.
$$

Finally, it is easy to check that $\sqrt{pq} \in [cp, q/c]$ for any $0<c \le \sqrt{q/p}$.

\subsection*{A9. Proof of Corollary~\ref{cr:Nash_confrontation}}
This is because:
\begin{equation}\label{eq:NBS_pi_ij}\tag{A.14}
	\underset{\lambda_{ji}>0}{\mathrm{argmax}} \left \{ \left (\lambda_{ji}-\frac{\pi_j}{\pi_i} \right ) \left (\lambda_{ij} - \frac{\pi_i}{\pi_j} \right ) \bigg| \lambda_{ij}\lambda_{ji}=1 \right \} = \frac{\pi_j}{\pi_i }
\end{equation}
and 
\begin{equation}\label{eq:NBS_g_ij}\tag{A.15}
	\underset{\lambda_{ji}>0}{\mathrm{argmax}} \left \{ \left (\lambda_{ji}-\frac{g_j}{g_i} \right ) \left (\lambda_{ij} - \frac{g_i}{g_j} \right ) \bigg| \lambda_{ij}\lambda_{ji}=1\right \} 
	= \frac{g_j }{g_i}.
\end{equation}
In Eqs.~(\ref{eq:NBS_pi_ij}) and (\ref{eq:NBS_g_ij}), both countries battle for better competitiveness and net trade balance, respectively.
At the solution points, the objective functions are zero and no country gets extra welfare,

\subsection*{A10. Proof of Corollary~\ref{cr:homogeneous_preference}}

When country $j$ directly competes with country $i$ to increase $\pi_j$, the Nash solution yields $\lambda_{ji}=\pi_j / \pi_i$. In this case, $\pi_j$ remains unchanged.
However, if country $j$ strategically aims to improve its trade balance instead, the Nash solution becomes $\lambda_{ji}=\sqrt{\pi_j g_j}/\sqrt{\pi_i g_i}$, which is typically larger than $\pi_j / \pi_i$ according to the Matthew effect.
Therefore, by Eq.~(\ref{eq:pi_j_2_P_ji}), country $j$ achieves its original objective of enhancing $\pi_j$, as its exports to $i$ increase --- regardless of a trade surplus or deficit.

Similarly, if country $j$ competes with $i$ solely to improve its trade balance, the Nash solution is $\lambda_{ji}=g_j/g_i$, which results in zero trade surplus. But if country $j$ strategically seeks to enhance $\pi_j$, the Nash solution again becomes $\lambda_{ji}=\sqrt{\pi_j g_j}/\sqrt{\pi_i g_i}$, which is generally smaller than $g_j / g_i$ due to the Matthew effect. Consequently, an increasing in $P_{ji}$ leads to a positive trade surplus, fulfilling country $j$'s original objective.

\subsection*{A11. Proof of Theorem~\ref{thm:globalization}}

We write $M = [\xi_1, \xi_2, \cdots, \xi_n]$ where $\xi_j$ is the $j$th column of $M$. 
Since $\sigma_{ji}$ has a zero sum for all $j$, $\xi_1+\xi_2+\cdots+\xi_n = \vec{0}_n$.
We divide Eq.~(\ref{eq:linear_approximation}) by $\Delta P_{ii}$ and let $\Delta P_{ii}\to 0$ to get:
\begin{equation}\label{eq:pi_I_P_pi_M}\tag{A.16}
\frac{\mathrm{d} \pi}{\mathrm{d} P_{ii}}[I_n-P]
=
\pi \frac{\mathrm{d} P}{\mathrm{d} P_{ii}} 
=
\pi M
=
\left (\pi \xi_1, \cdots, \pi \xi_n \right ). 
\end{equation}
With the partition Eq.~(\ref{eq:partition_of_P}) of $P^\top$, the augmented matrix for the identity $\frac{\mathrm{d} \pi}{\mathrm{d} P_{ii}} \vec{1}_n = 0$ and the transpose of Eq.~(\ref{eq:pi_I_P_pi_M}) is:
\begin{equation}\label{eq:augment_ij_2}\tag{A.17}
\resizebox{.85\hsize}{!}{$
\left [
\begin{array}{ccccc|c}
\vec{1}_{i-1}^\top&1               &\vec{1}_{j-i-1}^\top&1               &\vec{1}_{n-j}^\top&0\\
I_{i-1}-H_1       &-\eta_1         & -H_2               &-\eta_2         &-H_3              &(\pi\xi_1,\cdots,\pi\xi_{i-1})^\top\\
-\mu_1            &\boxed{1-P_{ii}}& -\mu_2             &\boxed{-P_{ji}} &-\mu_3            &\pi \xi_i\\
-H_4              &-\eta_3         & I_{j-i-1}-H_5      &-\eta_4         &-H_6              &(\pi\xi_{i+1},\cdots,\pi\xi_{j-1})^\top\\
-\mu_4            &\boxed{-P_{ij}} & -\mu_5             &\boxed{1-P_{jj}}&-\mu_6            &\pi \xi_j\\
-H_7              &-\eta_5         & -H_8               &-\eta_6         &I_{n-j}-H_9       &(\pi\xi_{j+1},\cdots,\pi\xi_n)^\top
\end{array}
\right ].
$}
\end{equation}

As the last column of Eq.~(\ref{eq:augment_ij_2}) sums to zero, i.e., $\sum\limits_{j=1}^n \pi\xi_j = \pi \sum\limits_{j=1}^n \xi_j =0$, we add all rows (except for the first) to the $(i+1)$st row to make the $(i+1)$st row a zero vector.
After dropping the $(i+1)$st row and moving the $i$th column to the first column without changing the order of other columns, 
we get the augmented matrix for $\left (\frac{\mathrm{d} \pi_i}{\mathrm{d} P_{ii}}, \frac{\mathrm{d} \pi_{-i}^\top}{\mathrm{d} P_{ii}} \right )^\top$:
\begin{equation}\label{eq:simplied_augmented2}\tag{A.18}
\left[ 
\begin{array}{cc | c}
1        & \vec{1}_{n-1}^\top& 0 \\
-\alpha_i& I_{n-1}-Z_i       & (\pi M_i)^\top 
\end{array}
\right].
\end{equation}
We multiply the matrix of Eq.~(\ref{eq:left_multiplication}) to the left side of Eq.~(\ref{eq:simplied_augmented2}) to get:
\begin{equation}\label{eq:trans_augmented2}\tag{A.19}
\resizebox{.85\hsize}{!}{$
\left[ 
\begin{array}{cc | c}
1+\vec{1}_{n-1}^\top (I_{n-1}-Z_i)^{-1} \alpha_i&\vec{0}_{n-1}^\top&-\vec{1}^\top_{n-1} (I_{n-1} -Z_i)^{-1} (\pi M_i)^\top\\
-\alpha_i                                       &I_{n-1} -Z_i      &(\pi M_i)^\top 
\end{array}
\right].
$}
\end{equation}
Therefore, by the first row of Eq.~(\ref{eq:trans_augmented2}): 
$$
\frac{\mathrm{d} \pi_i}{\mathrm{d} P_{ii}} 
= 
-\frac{\vec{1}^\top_{n-1} (I_{n-1} -Z_i)^{-1} (\pi M_i)^\top}{1+\vec{1}_{n-1}^\top (I_{n-1}-Z_i)^{-1} \alpha_i}.
$$
Also, by the second row of Eq.~(\ref{eq:trans_augmented2}):
$$
-\frac{\mathrm{d} \pi_i}{\mathrm{d} P_{ii}} \alpha_i + \left (I_{n-1}-Z_i \right ) \frac{\mathrm{d} \pi_{_{-i}}}{\mathrm{d} P_{ii}} 
= 
(\pi M_i)^\top
$$
and thus:
$$
\begin{array}{rcl}
\frac{\mathrm{d} \pi_{_{-i}}}{\mathrm{d} P_{ii}}
&=&
\left (I_{n-1}-Z_i \right )^{-1} \left [
(\pi M_i)^\top
+ 
\frac{\mathrm{d} \pi_i}{\mathrm{d} P_{ii}} \alpha_i  
\right ] \\
&=&
\left(I_{n-1}-Z_i \right)^{-1} \left [ 
(\pi M_i)^\top
-
\frac{\vec{1}^\top_{n-1} (I_{n-1} -Z_i)^{-1} \left(\pi M_i\right)^\top}{1+\vec{1}_{n-1}^\top (I_{n-1}-Z_i)^{-1} \alpha_i}
\alpha_i
\right ].
\end{array}
$$

\subsection*{A12. Proof of Corollary~\ref{cr:zero_pi_pii}}

If $\pi M_i=\vec{0}_{n-1}^\top$, then $\frac{\mathrm{d} \pi_i}{\mathrm{d} P_{ii}}=0$ and $\frac{\mathrm{d} \pi_{-i}}{\mathrm{d} P_{ii}}=\vec{0}_{n-1}$ according to Eqs.~(\ref{eq:pi_i_2_P_ii}) and (\ref{eq:pi_minus_i_2_P_ii}). 
Thus, $\frac{\mathrm{d} \pi}{\mathrm{d} P_{ii}}=\vec{0}_n^\top$.

Conversely, if $\frac{\mathrm{d} \pi}{\mathrm{d} P_{ii}} = \vec{0}_n^\top$, then $\frac{\mathrm{d} \pi_i}{\mathrm{d} P_{ii}}=0$ and $\frac{\mathrm{d} \pi_{-i}}{\mathrm{d} P_{ii}}=\vec{0}_{n-1}$. 
Applying $\frac{\mathrm{d} \pi_i}{\mathrm{d} P_{ii}}=0$ to Eq.~(\ref{eq:pi_i_2_P_ii}), we have: 
\begin{equation}\label{eq:1_I_Z_Mi}\tag{A.20}
\vec{1}^\top_{n-1} (I_{n-1} -Z_i)^{-1} (\pi M_i)^\top=0.
\end{equation}
We plug Eq.~(\ref{eq:1_I_Z_Mi}) into Eq.~(\ref{eq:pi_minus_i_2_P_ii}) and use $\frac{\mathrm{d} \pi_{-i}}{\mathrm{d} P_{ii}}=\vec{0}_{n-1}$ to get:
$$
\left(I_{n-1}-Z_i \right)^{-1} \left(\pi M_i\right)^\top = \vec{0}_{n-1}.
$$
Therefore, $\left(\pi M_i\right)^\top = \left(I_{n-1}-Z_i \right) \vec{0}_{n-1}= \vec{0}_{n-1}$ and $\pi M_i = \vec{0}_{n-1}^\top.$

\subsection*{A13. Proof of Corollary~\ref{cr:zero_pi_pii_all}}
\noindent Since
$$
\resizebox{.90\hsize}{!}{$
\pi \mathrm{diag} \left(\sigma_{ii} \odot \Lambda_i\right)
=
\frac{-\left(\pi_1\lambda_{i1}P_{i1},\cdots,\pi_{i-1}\lambda_{i,i-1}P_{i,i-1},-\pi_i(1-P_{ii}),\pi_{i+1}\lambda_{i,i+1}P_{i,i+1},\cdots,\pi_n\lambda_{in}P_{in}\right)}{1-P_{ii}}
$},
$$
$\pi M_i = \vec{0}_{n-1}^\top$ implies that $-(1-P_{ii}) \pi M$, i.e.,
$$
\resizebox{.98\hsize}{!}{$
\left(\pi_1\lambda_{i1}P_{i1},\cdots,\pi_{i-1}\lambda_{i,i-1}P_{i,i-1},-\pi_i(1-P_{ii}),\pi_{i+1}\lambda_{i,i+1}P_{i,i+1},\cdots,\pi_n\lambda_{in}P_{in}\right)
\left[\begin{array}{c}\sigma_{1i} \\ \vdots \\ \sigma_{ni} \end{array} \right],
$}
$$
is a zero vector except for its $i$th element.

For any $t\neq i$, zero value of the $t$th element of the above vector implies that:
$$
\sum\limits_{j \neq i}\frac{\lambda_{ij}P_{ij}P_{jt}}{1-P_{ji}} \pi_j - \frac{\pi_i (1-P_{ii})P_{it}}{1-P_{ii}}=0.
$$ 
We sum the above equations over all $t\neq i$ to get:
$$
\sum_{t\neq i}\sum\limits_{j \neq i}\frac{\lambda_{ij}P_{ij}P_{jt}}{1-P_{ji}} \pi_j - \pi_i \sum_{t\neq i} P_{it}=0
$$
which leads to:
\begin{equation}\label{eq:special_case_lambda_ij}\tag{A.21}
\sum\limits_{j \in \mathcal{N}} \lambda_{ij}P_{ij} \pi_j = \pi_i.
\end{equation}
When $i$ runs all numbers in $\mathcal{N}$, thus, $\pi \left[ \Lambda \odot P \right]^\top  = \pi$.

Finally, if $\lambda_{ij}=\pi_i/\pi_j$, then Eq.~(\ref{eq:special_case_lambda_ij}) is clearly valid and so is Eq.~(\ref{eq:steady_state_pi_lambda}).
If $\lambda_{ij}=P_{ji}/P_{ij}$, then $\left[ \Lambda \odot P \right]^\top=P$ and, therefore, Eq.~(\ref{eq:steady_state_pi_lambda}) also holds.


\begin{thebibliography}{99}
	
\bibitem[Allen, 2000]{Allen2000}
Allen, B., 2000.
The future of microeconomic theory.
J. Econ. Persp. 14, 143--150.
\url{https://doi.org/10.1257/jep.14.1.143}.
	
\bibitem[Allen, Arkolakis, and Takahashi (2020)]{Allen&Arkolakis&Takahashi2020}
Allen, T., Arkolakis, C., Takahashi, Y., 2020.
Universal gravity.
J. Polit. Econ. 128, 393--433.
\url{https://doi.org/10.1086/704385}.

\bibitem[Anbarci and Sun, 2013]{Anbarci&Sun2013}
Anbarci, N., Sun, C., 2013.
Asymmetric Nash bargaining solutions: A simple Nash program.
Econ. Letters, 120(2), 211--214.
\url{https://doi.org/10.1016/j.econlet.2013.04.026}.
	
\bibitem[Bagwell, Staiger, and Yurukoglu (2020)]{Bagwell&Staiger&Yurukoglu2020}
Bagwell, K., Staiger, R.W., Yurukoglu, A., 2020.
`Nash-in-Nash' tariff bargaining.
J. Int. Econ. 122, 103--263.
\url{https://doi.org/10.1016/j.jinteco.2019.103263}.

\bibitem[Barney and Felin, 2013]{Barney&Felin2013}	
Barney, J., Felin, T., 2013.
What are microfoundations.
Academy Manag. Persp. 27, 138--155.
\url{https://doi.org/10.5465/amp.2012.0107}.

\bibitem[Bonacich, 1987]{Bonacich1987}
Bonacich, P., 1987. 
Power and centrality: a family of measures.
Amer. J. Sociol. 92, 1170--1182.
\url{https://doi.org/10.1086/228631}.

\bibitem[Chaney, 2014]{Chaney2014}
Chaney, T., 2014. 
The network structure of international trade.
Amer. Econ. Rev. 104, 3600--3634.
\url{https://doi.org/10.1257/aer.104.11.3600}.

\bibitem[Chung, Lee, and Osang (2016)]{Chung&Lee&Osang2016}
Chung, S., Lee, J., Osang, T., 2016.
Did China tire safeguard save U.S. workers?
Euro. Econ. Rev. 85, 22--38. 
\url{https://doi.org/10.1016/j.euroecorev.2015.12.009}.	
	
\bibitem[Copeland (2014)]{Copeland2014}
Copeland, D.C., 2014.
Economic Interdependence and War.
Princeton University Press, Princeton, NJ.
	
\bibitem[Delgado et al. (2012)]{Delgado&Ketels&Porter&Stern2012}
Delgado, M., Ketels, C., Porter, M.E., Stern, S., 2012. 
The determinants of national competitiveness.
NBER Working Paper 18249.
	
\bibitem[Gopinath et al., 2024]{Gopinath&Gourinchas&Presbitero&Topalova2024}
Gopinath, G., Gourinchas, P., Presbitero, A.F., Topalova, P., 2024.
Changing global linkages: a new cold war?
IMF Working Paper No. 2024/076, Washington, DC.
\url{https://doi.org/10.5089/9798400272745.001}.

\bibitem[Grossman and Helpman, 1995]{Grossman&Helpman1995}
Grossman, G.M., Helpman, E., 1995.  
Trade wars and trade talks.
J. Polit. Econ. 103, 675--708.
\url{https://doi.org/10.1086/261999}.
	
\bibitem[Hammond, 1987]{Hammond1987}
Hammond, P.J., 1987. 
On reconciling Arrow's theory of social choice with Harsanyi's fundamental utilitarianism, 
in: Feiwel, G.R. (Ed.), Arrow and the Foundations of the Theory of Economic Policy. 
Palgrave Macmillan, London, pp. 179--221.
\url{https://doi.org/10.1007/978-1-349-07357-3_5}.

\bibitem[Harrison and Rutstrom (1991)]{Harrison&Rutstrom1991}
Harrison, G.W., Rutstr$\mathrm{\ddot{o}}$m, E.E., 1991.
Trade wars, trade negotiations and applied game theory.
Econ. J. 101, 420--435.
\url{https://doi.org/10.2307/2233549}.

\bibitem[Harsanyi, 1955]{Harsanyi1955}
Harsanyi, J.C., 1955.
Cardinal welfare, individualistic ethics, and interpersonal comparisons of utility.
J. Polit. Econ. 63, 309--321.
\url{https://doi.org/10.1086/257678}.

\bibitem[Hu, 2020]{Hu2020}
Hu, X., 2020.
Sorting big data by revealed preference with application to college ranking.
J. Big Data 7, 30.
\url{https://doi.org/10.1186/s40537-020-00300-1}.
	
\bibitem[Hu and Shapley, 2003]{Hu&Shapley2003} 
Hu, X., Shapley, L.S., 2003.
On authority distributions in organizations: equilibrium.
Games Econ. Behav. 45, 132--152.
\url{https://doi.org/10.1016/S0899-8256(03)00130-1}.	
	
\bibitem[Isard (1954)]{Isard1954}
Isard, W., 1954. 
Location theory and trade theory: short-run analysis.
Quart. J. Econ. 68, 305--320.
\url{https://doi.org/10.2307/1884452}.
	
\bibitem[Karlin and Taylor (2012)]{Karlin&Taylor2012}
Karlin, S., Taylor, H.M., 2012.
A First Course in Stochastic Processes, second ed.
Academic Press, Cambridge, MA.

\bibitem[Lau et al., 2022]{Lau&Pal&Mahalik&Gozgor2022}
Lau, C. K., Pal, S., Mahalik, M. K., Gozgor, G., 2022. 
Economic globalization convergence in high and low globalized developing economies: implications for the post Covid-19 era. 
Econ. Anal. Policy 76, 1027--1039.
\url{https://doi.org/10.1016/j.eap.2022.10.013}.

\bibitem[McCallum (1995)]{McCallum1995}
McCallum, J., 1995.
National borders matter: Canada-U.S. regional trade patterns.
Amer. Econ. Rev. 85, 615--623.
\url{http://www.jstor.org/stable/2118191}.	
	
\bibitem[Nash, 1950]{Nash1950}
Nash, J., 1950.
The bargaining problem.
Econometrica 18, 155--162.
\url{https://doi.org/10.2307/1907266}.
	
\bibitem[New York Times, 2022]{NYT2022}
New York Times, October 7, 2022.
Biden Administration Clamps Down on China’s Access to Chip Technology.

\bibitem[Önder and Yilmazkuday, 2016]{Önder&Yilmazkuday2016}
$\ddot{\mathrm{O}}$nder, A.S., Yilmazkuday, H., 2016.
Trade partner diversification and growth: How trade links matter.
J. Macroeconomics 50 (2016): 241--258.

\bibitem[Organski, 1968]{Organski1968}	
Organski, A.F.K., 1968. 
World Politics, second ed. 
Alfred A. Knopf, New York.
	
\bibitem[Ossa, 2014]{Ossa2014} 
Ossa, R., 2014. 
Trade wars and trade talks with data.
Amer. Econ. Rev. 104, 4104--4146.
\url{https://doi.org/10.1257/aer.104.12.4104}.
	
\bibitem[Owen, 1972]{Owen1972}
Owen, G., 1972. 
Multilinear extensions of games.
Manag. Sci. 18, 64--79.
\url{https://doi.org/10.1287/mnsc.18.5.64}.
	
\bibitem[Porter (1985)]{Porter1985}
Porter, M.E., 1985.
Competitive Advantage: Creating and Sustaining Superior Performance.
Free Press, Washington, DC.
	
\bibitem[Pozsar, 2022]{Pozsar2022}
Pozsar, Z., 2022.
War and Industrial Policy.
Credit Suisse Economics.
\url{https://www.interest.co.nz/sites/default/files/2022-09/War%20%26%20industrial%20policy.pdf}.
	
\bibitem[Read, 2005]{Read2005}
Read, R., 2005. 
The political economy of trade protection: the determinants and welfare impact of the 2002 US Emergency Steel Safeguard Measures.
World Economy 28, 1119--1137. 
\url{https://doi.org/10.1111/j.1467-9701.2005.00722.x}.

\bibitem[Romer, 2015]{Romer2015}
Romer, P.M., 2015.
Mathiness in the theory of economic growth.
Amer. Econ. Rev. Papers \& Proc. 105, 89--93.
\url{https://doi.org/10.1257/aer.p20151066}.
	
\bibitem[Roth (1977)]{Roth1977}
Roth, A.E., 1977.
The Shapley value as a von Neumann-Morgenstern utility.
Econometrica 45, 657--664.
\url{https://doi.org/10.2307/1911680}.
	
\bibitem[Schneider (2005)]{Schneider2005}
Schneider, G., 2005. 
Capacity and concessions: bargaining power in multilateral negotiations.
J. Int. Stud. 33, 665--689.
\url{https://doi.org/10.1177/03058298050330031901}.
	
\bibitem[Shapley, 1953]{Shapley1953}
Shapley, L.S., 1953. 
A value for n-person games,
in: Kuhn, H., Tucker, A. (Eds.), Contributions to the Theory of Games II.
Princeton University Press, Princeton, NJ, pp.307--317.
\url{https://doi.org/10.1017/CBO9780511528446.003}.

\bibitem[Shapley and Shubik, 1954]{Shapley&Shubik1954}
Shapley, L. S., Shubik, M. 1954. 
A method for evaluating the distribution of power in a committee system. 
Amer. Polit. Sci. Rev. 48, 787--792. 
\url{https://doi.org/10.2307/1951053}.
		
\bibitem[Stiglitz (2002)]{Stiglitz2002}
Stiglitz, J.E., 2002. 
Globalization and Its Discontents.
W.W. Norton \& Company, New York.
	
\bibitem[Stiglitz (2017)]{Stiglitz2017}
Stiglitz, J.E., 2017. 
Globalization and Its Discontents Revisited: Anti-Globalization in the Era of Trump.
W.W. Norton \& Company, New York, NY.
	
\bibitem[Stutz and Warf, 2010]{Stutz&Warf2010}
Stutz, F.P., Warf, B., 2010. 
The World Economy: Resources, Location, Trade, and Development.
Prentice Hall, Englewood Cliffs, NJ.
	
\bibitem[United Nations (2021)]{UN2021}
United Nations ComTrade Databases. 
\url{https://comtrade.un.org/}. Accessed 1 October 2021.
	
\bibitem[von Neumann-Morgenstern (1953)]{vonNeumann&Morgenstern1953}
Von Neumann, J., Morgenstern, O., 1953.
Theory of Games and Economic Behavior, third ed.
Princeton University Press, Princeton, NJ.
\url{https://doi.org/10.1515/9781400829460}.

\bibitem[Washington Post, 2018]{WashingtonPost2018a}
Washington Post, April 4, 2018a. 
China fires back at Trump with the threat of tariffs on 106 U.S. products, including soybeans. 

\bibitem[Washington Post(2018)]{WashingtonPost2018b}
The Washington Post, March 1, 2018b.
Trump announces steel and aluminum tariffs Thursday over objections from advisers and Republicans.

\bibitem[WITS Database (2021)]{WITS2021}
World Integrated Trade Solutions (WITS) of World Bank. 
\url{https://wits.worldbank.org/}. Accessed 1 October 2021.
	
\bibitem[Yan, 2014]{Yan2014}
Yan, X., 2014.
From keeping a low profile to striving for achievement.
Chinese J. Int. Polit. 7, 153--184.
\url{https://doi.org/10.1093/cjip/pou027}.

\bibitem[Yilmazkuday, 2021]{Yilmazkuday2021}
Yilmazkuday, H., 2021.
Drivers of Global Trade: A Product-Level Investigation.
Intl. Econ. J. 35, 469--485.
\url{https://doi.org/10.1080/10168737.2021.1983631}.

\bibitem[Yilmazkuday and Yilmazkuday, 2014]{Yilmazkuday&Yilmazkuday2014}
Yilmazkuday, D., Yilmazkuday, H., 2014.
Bilateral versus multilateral free trade agreements: A welfare analysis. 
Rev. Intl. Econ. 22, 513--535.
\url{https://doi.org/10.1111/roie.12131}.

\end{thebibliography}
\end{document}